\documentclass[longbibliography,10pt,aps,prx,twocolumn,notitlepage,superscriptaddress]{revtex4-1}
\usepackage{amsthm,amsfonts,verbatim,color}
\usepackage{amsmath,amssymb}
\usepackage{graphicx}
\usepackage[export]{adjustbox}
\usepackage[caption=false]{subfig}
\usepackage{bm}
\usepackage{epsfig,slashed}
\usepackage{amssymb}
\usepackage{amsfonts}
\usepackage{tikz}
\usepackage{xparse}
\usepackage{ifthen}
\usepackage{lipsum}
\usepackage{siunitx}

\usepackage{bbold}  
\DeclareMathAlphabet{\mathbbold}{U}{bbold}{m}{n}

\usepackage[colorlinks=true,citecolor=cyan,linkcolor=magenta,filecolor=magenta]{hyperref}
\usepackage{orcidlink}

\begin{document}

\newcommand{\tr}{\mathop{\mathrm{tr}}}
\newcommand{\bsigma}{\boldsymbol{\sigma}}
\newcommand{\bphi}{\boldsymbol{\phi}}
\renewcommand{\Re}{\mathop{\mathrm{Re}}}
\renewcommand{\Im}{\mathop{\mathrm{Im}}}
\renewcommand{\b}[1]{{\boldsymbol{#1}}}
\newcommand{\diag}{\mathrm{diag}}
\newcommand{\sign}{\mathrm{sign}}
\newcommand{\sgn}{\mathop{\mathrm{sgn}}}
\renewcommand{\c}[1]{\mathcal{#1}}
\newcommand{\Jac}{\mathop{\mathrm{Jac}}}
\newcommand{\Mob}{\mathop{\textrm{M\"ob}}}
\newcommand{\Aut}{\mathop{\mathrm{Aut}}}
\renewcommand{\g}{\mathfrak{g}}
\renewcommand{\geq}{\geqslant}
\renewcommand{\leq}{\leqslant}
\newcommand{\PBC}{\text{PBC}}
\def\abs#1{\left|{#1}\right|} 

\newcommand{\halfs}{\mbox{\small{$\frac{1}{2}$}}} 
\newcommand{\Nf}{N_{\!f}}
\newcommand{\partialslash}{\partial \! \! \! /}
\newcommand{\xslash}{x \! \! \! /}
\newcommand{\yslash}{y \! \! \! /}

\newcommand{\cl}{\mathrm{cl}}
\newcommand{\mb}{\bm}
\newcommand{\ua}{\uparrow}
\newcommand{\da}{\downarrow}
\newcommand{\ra}{\rightarrow}
\newcommand{\la}{\leftarrow}
\newcommand{\mc}{\mathcal}
\newcommand{\bs}{\boldsymbol}
\newcommand{\lra}{\leftrightarrow}
\newcommand{\nn}{\nonumber}
\newcommand{\half}{{\textstyle{\frac{1}{2}}}}
\newcommand{\mf}{\mathfrak}
\newcommand{\MF}{\text{MF}}
\newcommand{\IR}{\text{IR}}
\newcommand{\UV}{\text{UV}}
\newcommand{\sech}{\mathrm{sech}}

\newcommand*{\vcenteredhbox}[1]{\begingroup
\setbox0=\hbox{#1}\parbox{\wd0}{\box0}\endgroup}
\newcommand{\picscalefactor}{0.5}

\newcommand{\addextralinespace}[1]{\rule[#1\normalbaselineskip]{0pt}{0pt}}
	

\definecolor{TB}{rgb}{1,0.5,0}
\def\TB#1{{\color{TB}#1}}
\def\TBC#1{{\color{TB}\xout{#1}}}
\def\TBD#1{{\color{TB}[TB: {#1}]}}

\definecolor{JM}{rgb}{0,0.5,1}
\def\JM#1{{\color{JM}#1}}
\def\JMC#1{{\color{JM}[JM: {#1}]}}

\title{Flat bands and band touching from real-space topology in hyperbolic lattices}

\author{Tom\'a\v{s} Bzdu\v{s}ek\,\orcidlink{0000-0001-6904-5264}}
\email{tomas.bzdusek@psi.ch}
\affiliation{Condensed Matter Theory Group, Paul Scherrer Institute, 5232 Villigen PSI, Switzerland}
\affiliation{Department of Physics, University of Zurich, Winterthurerstrasse 190, 8057 Zurich, Switzerland}
\author{Joseph Maciejko\,\orcidlink{0000-0002-6946-1492}}
\email{maciejko@ualberta.ca}
\affiliation{Department of Physics \& Theoretical Physics Institute (TPI), University of Alberta, Edmonton, Alberta T6G 2E1, Canada}

\date{\today}

\begin{abstract}
Motivated by the recent experimental realizations of hyperbolic lattices in circuit quantum electrodynamics and in classical electric-circuit networks, we study flat bands and band-touching phenomena in such lattices. We analyze noninteracting nearest-neighbor hopping models on hyperbolic analogs of the kagome and dice lattices with heptagonal and octagonal symmetry. 
We show that two characteristic features of the energy spectrum of those models, namely the fraction of states in the flat band as well as the number of touching points between the flat band and the dispersive bands, can both be captured exactly by a combination of real-space topology arguments and a reciprocal-space description via the formalism of hyperbolic band theory. Furthermore, using real-space numerical diagonalization on finite lattices with periodic boundary conditions, we obtain new insights into higher-dimensional irreducible representations of the non-Euclidean (Fuchsian) translation group of hyperbolic lattices. 
First, we find that the fraction of states in the flat band is the same for Abelian and non-Abelian hyperbolic Bloch states. 
Second, we find that only Abelian states participate in the formation of touching points between the flat and dispersive bands.
\end{abstract}

\maketitle

\section{Introduction}

Recent years have witnessed much interest in condensed matter systems that exhibit {\it flat bands}, i.e., energy bands of a tight-binding Hamiltonian that are independent of the crystal momentum~\cite{leykam2018,rhim2021}. Since the ratio of interaction potential energy to kinetic energy diverges for a dispersionless band, flat-band systems are fertile grounds to engineer strongly correlated states~\cite{derzhko2015}, including ferromagnetism~\cite{tasaki1998}, superconductivity~\cite{aoki2020}, Wigner crystallization~\cite{wu2007}, and the fractional quantum Hall effect~\cite{tang2011,sun2011,Neupert:2011,sheng2011,regnault2011,parameswaran2013}. For example, Moir\'e superlattices~\cite{bistritzer2011} support nearly-flat bands with a nontrivial fragile band topology~\cite{zou2018,po2018} that are believed to be key to understanding the exotic correlated states found in twisted bilayer graphene~\cite{cao2018,cao2018b}.

In the noninteracting limit, key properties of flat bands such as band degeneracy and possible touchings with other dispersive bands can be understood from an intriguing interplay between \emph{real-space} and \emph{reciprocal-space} physics~\cite{Bergman:2008,rhim2019}. 
Recall that for conventional (Euclidean) condensed matter lattices in two dimensions (2D), the reciprocal space takes the form of a 2D torus parametrized with the crystal momentum vector ($\bs{k}$)---allowing us to equivalently call it the \emph{momentum space}.
In the absence of topological obstructions, the (symmetry-compatible and exponentially localized) Wannier functions associated to an \emph{isolated flat band} can be constructed from a set of compact 
localized states (CLS), so named because they have nonvanishing support on a finite number of sites~\cite{Sutherland:1986,lieb1989,flach2014,maimaiti2017,read2017,ramachandran2017,graf2021}. 
On the other hand, band touchings between a flat band and a dispersive band can be shown to arise from extended eigenstates, whose support extends along noncontractible loops in real space for a toroidal sample with periodic boundary conditions (PBC)~\cite{Bergman:2008}. 
Such band touchings are thus topologically protected, but by topology in real space rather than in momentum space. 
While of a topological kind, the arguments of Ref.~\onlinecite{Bergman:2008} rely crucially on the Euclidean geometry of conventional crystalline lattices, which ensures the toroidal nature of both real space and momentum space under PBC (as well as the very existence of momentum space via Bloch's theorem).

In this work, we study flat bands and band-touching phenomena in \emph{hyperbolic lattices}. 
With recent experimental realizations in circuit quantum electrodynamics (cQED)~\cite{Kollar:2019} and electrical circuits~\cite{Lenggenhager:2021}, hyperbolic lattices are periodic in the non-Euclidean sense and correspond to regular tessellations of the \emph{hyperbolic plane}~\cite{coxeter1957}, i.e., a 2D space of uniform negative curvature~\cite{balazs1986}. 
Hyperbolic lattices have recently become a fertile ground to investigate interplay of the negative curvature with a variety of physics phenomena, including the Bloch theorem~\cite{Maciejko:2021,Maciejko:2022,cheng2022}, Hofstadter spectra~\cite{yu2020,ikeda2021,stegmaier2021}, topological phases~\cite{Urwyler:2021,urwyler2022,liu2022,zhang2022,chen2022}, and strong correlations~\cite{daniska2016,zhu2021,bienias2022,Boettcher:2020}. Notably, in the cQED experiment~\cite{Kollar:2019}, the nearest-neighbor tight-binding model on the so-called heptagon-kagome lattice (Fig.~\ref{fig:7kagome-sites}) was simulated. 
As demonstrated by numerical diagonalization on finite lattices with open boundary conditions (OBC), the model possesses a flat band with macroscopic degeneracy, which is separated from the rest of the spectrum by a gap~\cite{Kollar:2020,kollar2021}.
As observed in Ref.~\onlinecite{Kollar:2019}, the real-space topology arguments of Ref.~\onlinecite{Bergman:2008} cannot be used to understand flat-band phenomena in this or other hyperbolic lattices, unless Bloch theory is generalized to such lattices.

Here, we combine real-space topology techniques in hyperbolic space, together with the newly developed \emph{hyperbolic band theory} (HBT)~\cite{Maciejko:2021,Maciejko:2022} and \emph{crystallography of hyperbolic lattices}~\cite{Boettcher:2021}, to understand flat bands and band-touching phenomena in hyperbolic lattices. 
HBT exploits the non-Euclidean translation symmetry of hyperbolic lattices, which is captured by a discrete non-Abelian group known as a \emph{Fuchsian translation group}, to develop a reciprocal-space description of such lattices. 
The familiar 2D Brillouin-zone torus is replaced by a collection of generalized Brillouin zones. 
These comprise a higher-dimensional torus of 1D irreducible representations (irreps) of the Fuchsian translation group, henceforth called the \emph{hyperbolic momentum space}, but also an infinite sequence of moduli spaces that parametrize higher-dimensional irreps of the Fuchsian translation group.
We specifically focus on four distinct lattices: the octagon-kagome (Fig.~\ref{fig:8kagome-sites}) and heptagon-kagome (Fig.~\ref{fig:7kagome-sites}) lattices, which generalize the kagome lattice, and the octagon-dice (Fig.~\ref{fig:8dice-sites}) and heptagon-dice (Fig.~\ref{fig:7dice-sites}) lattices, which generalize the dice lattice~\cite{Sutherland:1986}. 
The octagonal lattices possess a flat band which touches a dispersive band, like their Euclidean counterparts~\cite{Bergman:2008}, while the flat bands of heptagonal lattices are gapped. 

\begin{table}[t]
\caption{Flat-band fraction $f$ and band-touching index $w$ for the hyperbolic lattices studied in this work [see Eq.~(\ref{FBtotal})].}
\begin{ruledtabular}
\begin{tabular}{ccc}
	Lattice         &   $f$     &	$w$ \\ \hline
    octagon-kagome  &   1/3     &	1	\\ 
    octagon-dice    &  5/11		&	2	\\ 
    heptagon-kagome &  1/3		&	0	\\ 
    heptagon-dice   &  2/5		&	0	\\ 
\end{tabular}
\end{ruledtabular}
\label{tab:summary}
\end{table}

The main result of our analysis can be summarized as follows. The key differences between Euclidean and non-Euclidean flat bands, as well as the precise fraction of states that lie in the flat band and the number of band-touching points (Table~\ref{tab:summary}), follow ultimately from the unusual higher-genus topology imposed by PBC in hyperbolic space~\cite{Maciejko:2022,sausset2007}, which is captured by our real-space and HBT arguments. 
The unique combination of real-space and (hyperbolic-)momentum-space characterization allows us to also capture certain concrete aspects of bulk eigenstates transforming according to non-Abelian irreps of the Fuchsian translation group~\cite{Maciejko:2022}---a feat not achieved by prior works utilizing HBT~\cite{urwyler2022,chen2022}. 
Namely, we find that the fraction of the spectral weight lying in the flat band is the same for both Abelian and non-Abelian irreps of the Fuchsian translation group. In addition, the states responsible for the touching of the flat band with the remainder of the spectrum are found to belong to Abelian irreps.

The rest of the paper is structured as follows. In Sec.~\ref{sec:gen}, we outline our general strategy to characterize the four chosen hyperbolic lattices. In short, we focus on finite lattices with PBC and use a combination of analytical real-space arguments, HBT calculations, and real-space numerical diagonalization to compute the flat-band fraction and the degeneracy of band-touching points for each lattice. 
The use of PBC ensures we are capturing properties of the bulk spectrum, eliminating any edge effects. In the four subsequent sections (Sec.~\ref{sec:8kag}-\ref{sec:7dice}), we apply this strategy to the octagon-kagome, octagon-dice, heptagon-kagome, and heptagon-dice lattices, respectively. Finally, in Sec.~\ref{sec:summary}, we summarize our results and outline directions for future research. 

\section{General strategy}
\label{sec:gen}

We consider hyperbolic analogs of two Euclidean lattices with uniform nearest-neighbor hopping that harbor flat energy bands, namely the \emph{kagome} and \emph{dice} lattices~\cite{Bergman:2008,Sutherland:1986}. 
We set the hopping amplitude to $t$.
Additionally, for the dice lattice we consider inclusion of an on-site potential $V_B$ to sites with coordination number larger than $3$. As will be seen, such an on-site potential preserves the symmetries of the dice lattice and serves to lift accidental degeneracies that are not protected by real-space topology~\cite{Bergman:2008}.
The flat-band energy is $E=-2t$ for the kagome lattice and $E=0$ for the dice lattice (for both the Euclidean and the hyperbolic versions).

The analysis of Ref.~\cite{Bergman:2008} crucially relies on the ability to decompose a finite lattice with PBC into $N$ unit cells related by translations; we will call such a finite lattice a \emph{($N$-cell) PBC cluster}.
As shown in Refs.~\cite{Maciejko:2021,Maciejko:2022,Boettcher:2021}, the proper generalization of unit cell for hyperbolic lattices is the fundamental domain of a strictly hyperbolic Fuchsian group $\Gamma$ (also called the Fuchsian translation group). 
The ``strictly hyperbolic'' condition means that $\Gamma$ only contains hyperbolic elements, which are boost-like isometries of hyperbolic space~\cite{Katok}. 
This infinite group maps each fundamental domain to another without fixed points and can thus be interpreted as a group of (non-Abelian) translations. 
In the following, we will use the simpler terminology of \emph{unit cell} to denote such a fundamental domain. 

The centers of such unit cells that tile the hyperbolic plane constitute a \emph{hyperbolic Bravais lattice}~\cite{Boettcher:2021}.
We represent hyperbolic lattices using the Poincar\'e-disk model of hyperbolic space~\cite{balazs1986}, in which the analog of a straight line (geodesic) is an arc of circle normal to the bounding circle. 
For the octagonal lattices, the unit cell is a regular hyperbolic octagon (red octagon in Figs.~\ref{fig:8kagome-sites} and~\ref{fig:8dice-sites}) which we will call the {\it Bolza cell}. 
Conversely, for the heptagonal lattices, the unit cell is a regular hyperbolic 14-sided polygon (red 14-gon in Figs.~\ref{fig:7kagome-sites} and~ \ref{fig:7dice-sites}) which we will call the {\it Klein cell}. 
These names refer to the fact that, by identifying pairwise sides of these hyperbolic polygons under the quotienting action of the Fuchsian group $\Gamma$, one obtains two compact Riemann surfaces known as the Bolza surface~\cite{bolza1887} and the Klein surface (or Klein quartic)~\cite{klein1878,EightfoldWay}, respectively. 
When viewed as complex manifolds, as opposed to purely topological surfaces, these Riemann surfaces possess the largest amount of symmetry available in their given genus (2 and 3, respectively)~\cite{Miranda,kazaryan2019}, which will aid in the systematic construction of CLS for the corresponding lattices later on.

We analyze each lattice using both a real-space perspective, and a reciprocal-space perspective via HBT. 
The structure of reciprocal space is more involved in HBT than in Euclidean band theory~\cite{Maciejko:2021,Maciejko:2022}. 
First, the ordinary 2D Brillouin zone torus becomes a $2g$-dimensional torus parametrized by the hyperbolic crystal momentum $\b{k}=(k_1,\ldots,k_{2g})$, where $g$ is the genus of the unit-cell Riemann surface. 
Thus, the Brillouin zone is 4D for the octagonal lattices, while it is 6D for the heptagonal lattices. 
Second, since the translation group $\Gamma$ is non-Abelian, it also admits higher-dimensional unitary irreps, in addition to the 1D irreps described by the 4D/6D Brillouin zones. 
These higher-dimensional irreps live in certain moduli spaces (also known as character varieties)~\cite{Maciejko:2022,kienzle2022} that can be viewed as non-Abelian Brillouin zones. 
To summarize, the hyperbolic lattices we study here support two types of Bloch states: those with crystal momentum $\b{k}$ in a 4D/6D Brillouin zone, referred to as {\it Abelian states}, and those corresponding to higher-dimensional irreps, dubbed {\it non-Abelian states}.

We can now formulate our strategy more precisely. For each lattice, we perform the following analysis:
\begin{enumerate}
    \item[{\bf A.}] {\bf Real space}: Given a PBC cluster of $N$ unit cells, we construct a complete basis of CLS and extended noncontractible-loop states that span the flat band.
    \item[{\bf B.}] {\bf Reciprocal space (HBT)}:
    \begin{enumerate}
        \item[1.] We construct a complete basis of Abelian flat-band states with $\b{k}=\b{0}$.
        \item[2.] We construct a complete basis of Abelian flat-band states with $\b{k}\neq\b{0}$.
        \item[3.] We study non-Abelian flat-band states using numerical exact diagonalization on PBC clusters.
    \end{enumerate}
\end{enumerate}
For each lattice, the results of our analysis can be summarized as follows. From \textbf{A} above, we find that the total number of eigenstates at the energy of the flat band, $N_\text{FBS}$, is related to the total number of eigenstates in the spectrum, $N_\text{S}\propto N$, by the following equation:
\begin{align}\label{FBtotal}
    N_\text{FBS}=f N_\text{S}+w,
\end{align}
where we call $f$ the {\it flat-band fraction} and $w$ the {\it band-touching index} (Table~\ref{tab:summary}). While $N_\text{FBS}$ and $N_\text{S}$ both grow with the system size $N$, $f$ and $w$ are pure numbers independent of $N$, which characterize a given lattice. From \textbf{B}, we find that Eq.~(\ref{FBtotal}) can be refined as follows:
\begin{align}
    N_\text{FBS}^\text{ab}&=f N^\text{ab}+w,\label{FBfrac1}\\
    N_\text{FBS}^\text{nonab}&=f N^\text{nonab},\label{FBfrac2}
\end{align}
where $N^\text{ab}$ and $N^\text{nonab}$ denote the total number of Abelian and non-Abelian states in the spectrum, respectively, and $N_\text{FBS}^\text{ab}$ and $N_\text{FBS}^\text{nonab}$ denote the number of such states in the flat band. Adding Eqs.~(\ref{FBfrac1}) and (\ref{FBfrac2}), we recover Eq.~(\ref{FBtotal}), since a given state can always be classified as either Abelian or non-Abelian ($N_\text{S}=N^\text{ab}+N^\text{nonab}$ and $N_\text{FBS}=N_\text{FBS}^\text{ab}+N_\text{FBS}^\text{nonab}$). 

The physical meaning of these equations is as follows. Equation~(\ref{FBfrac1}) follows from comparing \textbf{B}.1 and \textbf{B}.2, and indicates an enhanced degeneracy of the flat band at $\b{k}=\b{0}$ compared to nonzero momenta $\b{k}\neq\b{0}$. In HBT, this corresponds to a band touching at the origin of the 4D/6D Brillouin zone between the flat band and $w$ dispersive bands. Equation~(\ref{FBtotal}), obtained via \textbf{A}, shows that this reciprocal-space phenomenon can be understood from a real-space topology argument, where extended noncontractible-loop states play a key role. Finally, Eq.~(\ref{FBfrac2}) is obtained by taking the difference of Eq.~(\ref{FBtotal}) and Eq.~(\ref{FBfrac1}), as well as corroborated numerically from \textbf{B}.3. This equation reveals that non-Abelian states do not participate in the band touching, but are characterized by the same flat-band fraction $f$ as the Abelian states. In cases where $w=0$, there is no band touching protected by real-space topology, and we correspondingly observe that the flat band is gapped.

When analyzing the spectrum from the real-space perspective (\textbf{A}), we attempt to construct flat-band CLS that are maximally localized, i.e., with support on the smallest possible number of sites. We define a measure of localization in terms of the underlying $\{p,3\}$ or dual $\{3,p\}$ lattices, with $p=8$ for the octagonal lattices and $p=7$ for the heptagonal lattices. (Here, the Schl\"{a}fli symbol $\{p,q\}$ indicates a 2D tessellation by regular $p$-sided polygons such that $q$ of them meet at each vertex.) For both octagonal lattices, the maximally localized CLS is bound to a single octagonal face of the $\{8,3\}$ lattice, or equivalently a single vertex of the dual $\{3,8\}$ lattice.
In contrast, the maximally localized CLS for the heptagon-kagome lattice is bound to a {\it pair} of heptagonal faces of the $\{7,3\}$ lattice~\cite{Kollar:2019,Kollar:2020}, or equivalently a single {\it edge} of the $\{3,7\}$ lattice. 
Finally, we find that the maximally localized CLS for the heptagon-dice lattice is bound to a {\it triplet} of heptagonal faces of the $\{7,3\}$ lattice, or equivalently a single triangular {\it face} of the $\{3,7\}$ lattice.

\section{Octagon-kagome lattice}
\label{sec:8kag}

\begin{figure}[t!]
\centering
    \includegraphics[width=0.855\linewidth]{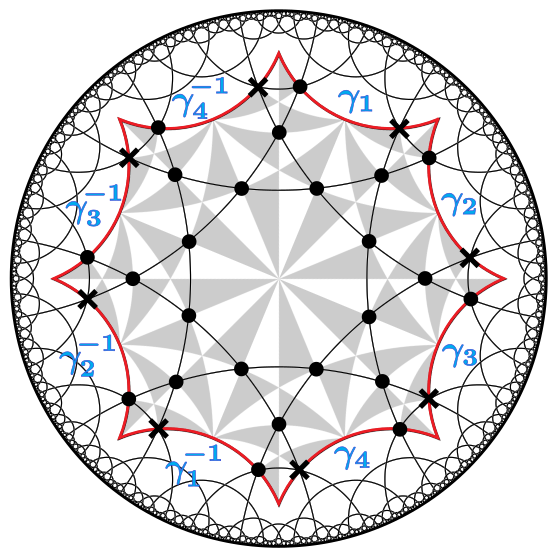}
    \caption[]{
    Octagon-kagome lattice. The 8-sided Bolza cell (red), which contains 24 inequivalent sites (black dots), can be translated by the Fuchsian group generators $\gamma_j$, $j=1,\ldots,4$ (blue) to generate the entire lattice. Crosses indicate sites on the cell boundary that are related by Fuchsian translations to sites that are already taken to belong to the displayed Bolza cell, i.e., the crossed sites belong to adjacent unit cells.
    The 96 grey/white triangles are fundamental domains for the action of the symmetry group, isomorphic to $\mathrm{GL}(2,\mathbb{Z}_3)\rtimes\mathbb{Z}_2$, of the Bolza surface.
    }
\label{fig:8kagome-sites}
\end{figure}

We now proceed to derive the results summarized in Table~\ref{tab:summary}, following the steps \textbf{A} and \textbf{B} outlined above, beginning with the octagon-kagome lattice (Fig.~\ref{fig:8kagome-sites}). The octagon-kagome lattice is the line graph of the $\{8,3\}$ lattice. In general, the line graph $L(X)$ of a graph $X$ is obtained by placing vertices on the edges of $X$, and connecting those vertices if the underlying edges share a common vertex in $X$. Hopping models on line graphs generally support flat bands~\cite{Kollar:2020,chiu2020,kollar2021}. Kagome-like lattices are line graphs of $\{p,3\}$ lattices, with $p=6$ for the ordinary (Euclidean) kagome lattice, $p=7$ for the heptagon-kagome lattice studied in Sec.~\ref{sec:7kag}, and $p=8$ for the octagon-kagome lattice studied in this section.

As mentioned previously, the unit cell of the octagon-kagome lattice is the Bolza cell, a regular 8-sided hyperbolic polygon. Translating the Bolza cell under the action of the Fuchsian group generated by four elementary translations $\gamma_j$, $j=1,\ldots,4$ and their inverses, one obtains a regular $\{8,8\}$ tessellation of the Poincar\'e disk~\cite{Maciejko:2021}. In other words, the hyperbolic Bravais lattice for the $\{8,3\}$ lattice and its octagon-kagome line graph is the $\{8,8\}$ lattice~\cite{Boettcher:2021}. Thus we denote the Fuchsian group as $\Gamma_{\{8,8\}}$, and its presentation is given as:
\begin{align}
    \Gamma_{\{8,8\}}=\langle\gamma_1,\gamma_2,\gamma_3,\gamma_4:\gamma_1\gamma_2^{\!-1}\gamma_3\gamma_4^{\!-1}\gamma_1^{\!-1}\gamma_2\gamma_3^{\!-1}\gamma_4\!=\!\mathbbold{1}\rangle,
\end{align}
where $\mathbbold{1}$ denotes the identity element. The Bolza cell contains 24 inequivalent vertices of the octagon-kagome lattice (black dots in Fig.~\ref{fig:8kagome-sites}); thus a finite PBC cluster with $N$ Bolza cells contains a total of $24N$ sites, and its hopping Hamiltonian possesses a total of $N_\text{S}=24N$ eigenstates.

To construct PBC clusters of the octagon-kagome lattice, we follow the approach of Ref.~\cite{Maciejko:2022}. A PBC cluster is associated to a normal subgroup $\Gamma_\text{PBC}$ of the Fuchsian group $\Gamma_{\{8,8\}}$, which is denoted by $\Gamma_\text{PBC}\triangleleft\Gamma_{\{8,8\}}$. For a PBC cluster with $N$ Bolza cells, this normal subgroup is of index $N$ in $\Gamma_{\{8,8\}}$, denoted as $|\Gamma_{\{8,8\}}:\Gamma_\text{PBC}|=N$. The index corresponds to the number of cosets of $\Gamma_\text{PBC}$ in $\Gamma_{\{8,8\}}$, i.e., the number of inequivalent unit cells modulo translations that span the entire cluster (elements of $\Gamma_\text{PBC}$). Physically, this describes a choice of $N$ adjacent Bolza cells (known in the mathematical literature as an $\{8,8\}$-animal~\cite{malen2021}) together with a choice of boundary identifications. For a given $N$, all possible normal subgroups of $\Gamma$ are constructed using an algorithm known as the low-index normal subgroups procedure~\cite{dietze1974,conder2005,FirthThesis}, implemented for the computer algebra package GAP~\cite{GAP4} by F.~Rober~\cite{LINS}. Topologically, a PBC cluster with $N$ Bolza cells is an $N$-fold cover of the Bolza surface; it is thus a surface of genus $h$ given by the Riemann-Hurwitz formula~\cite{Miranda}: 
\begin{equation}
    h=N(g-1)+1.\label{eqn:Bolza-cluster-genus}
\end{equation}
Substituting $g=2$ for the genus of the Bolza surface, we find that $h=N+1$ for PBC clusters with $N$ Bolza cells.

\subsection{Flat bands from real-space topology}

\begin{figure}[t!]
\centering
    \includegraphics[width=0.855\linewidth]{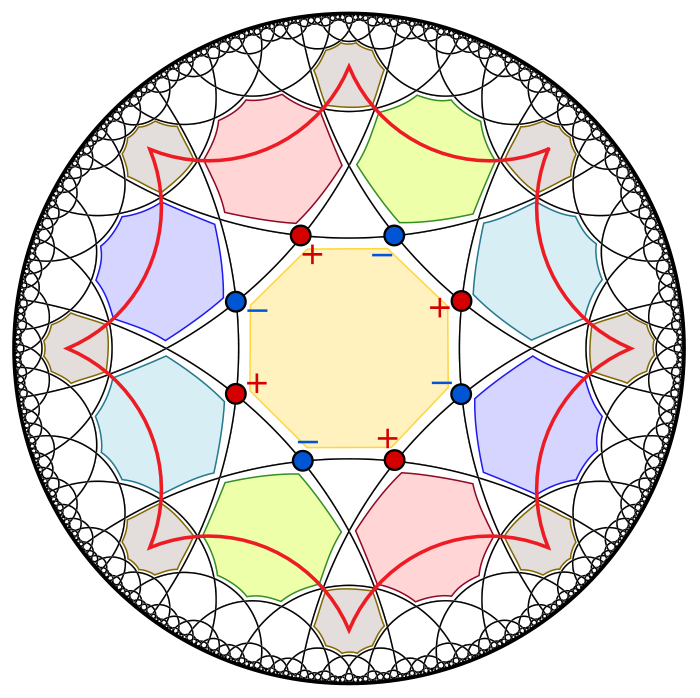}
    \caption[]{
    The maximally localized CLS on the octagon-kagome lattice are single-octagon states, with nonvanishing amplitude that alternates between $+1$ and $-1$ (red/blue dots) on the eight sites of a single octagon of the lattice. The six different colors represent the six inequivalent single-octagon states belonging to one Bolza cell.
    }
\label{fig:8kagome-CLS}
\end{figure}

The basic principle underlying flat-band states on kagome-like lattices is destructive interference on a single kagome triangle: having amplitudes $+1$ and $-1$ on two sites of such a triangle (e.g., red and blue dots in Fig.~\ref{fig:8kagome-CLS}) ensures that the probability of tunneling onto the third site is zero. (Here and elsewhere in the paper, we will ignore an overall normalization factor when quoting wave-function amplitudes.)

In the Euclidean kagome lattice, the maximally localized CLS which respects this principle is a state with support on a single hexagon, whose amplitude alternates between $+1$ and $-1$ as one goes around the hexagon~\cite{Bergman:2008}. Likewise here, the maximally localized CLS with energy $E=-2t$ is the {\it single-octagon state}, whose amplitude alternates between $+1$ and $-1$ on the sites of a single octagon of the octagon-kagome lattice. As illustrated in Fig.~\ref{fig:8kagome-CLS}, there are six inequivalent single-octagon states in each Bolza cell; this leads to a total of $6N$ single-octagon states for the entire PBC cluster. On a closed surface with PBC, there is one constraint on the linear independence of those $6N$ states: namely, the equal-weight superposition of all those states vanishes. Indeed, each site of the octagon-kagome lattice shares two single-octagon states whose amplitudes on that site can be made equal and opposite. As a result, there are $6N-1$ linearly independent CLS on a PBC cluster with $N$ Bolza cells.

\begin{figure}[t!]
\centering
    \includegraphics[width=0.855\linewidth]{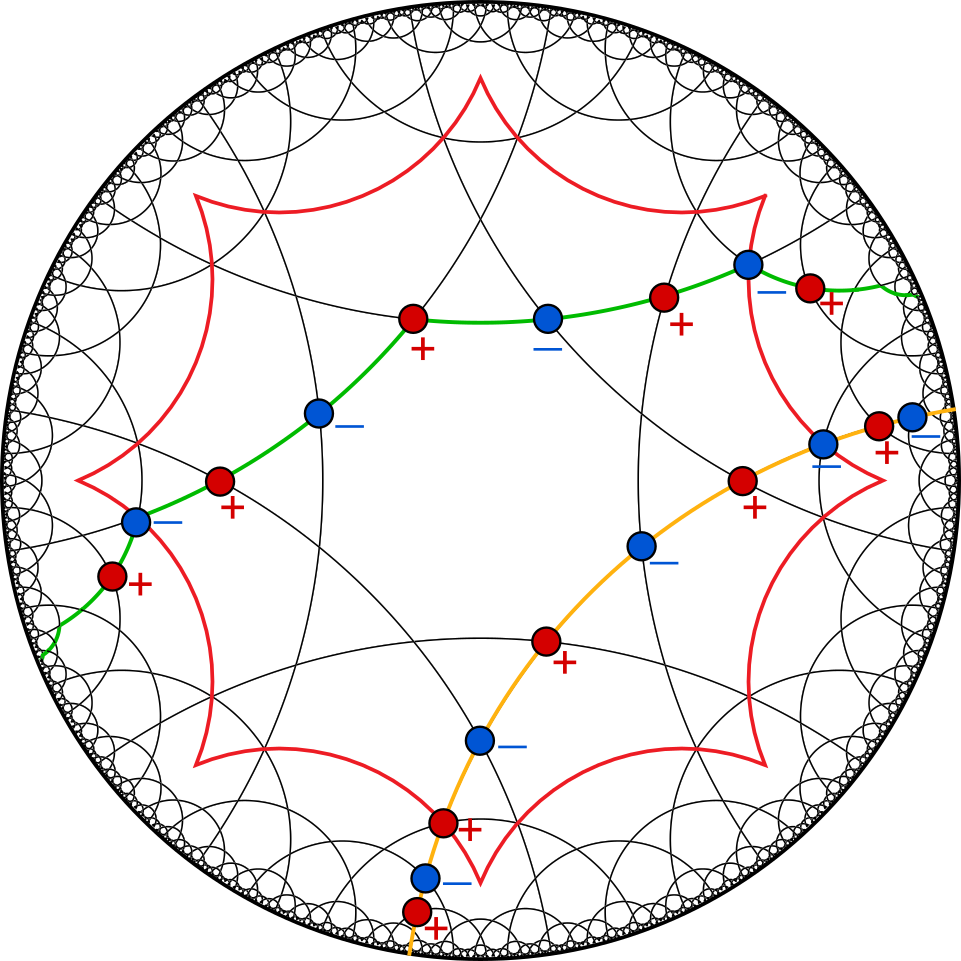}
    \caption[]{
    Two examples of flat-band states that extend along noncontractible loops on PBC clusters of the octagon-kagome lattice.
    The state extending along the green path is compatible with PBC on a single Bolza cell (see Fig.~\ref{fig:8kagome-sites}) and corresponds to one of the 4 noncontractible-loop states discussed in Sec.~\ref{sec:8kag-HBT-1}. The state on the orange path extends along a hyperbolic geodesic.
    }
\label{fig:8kagome-homology}
\end{figure}

As for the Euclidean kagome lattice, however, there exist flat-band states on the octagon-kagome lattice that cannot be captured by the $6N-1$ CLS. These states have support along an extended chain of sites that winds nontrivially along noncontractible loops in a compactified PBC cluster (Fig.~\ref{fig:8kagome-homology}). Single-octagon states and their linear combinations can only form contractible loops, thus the noncontractible-loop states are necessarily linearly independent from the CLS of Fig.~\ref{fig:8kagome-CLS}. A compact surface of genus $h$ possesses $2h$ noncontractible loops~\cite{Nakahara}. Since a PBC cluster with $N$ Bolza cells is a compact surface of genus $N+1$, as mentioned earlier, there are $2N+2$ noncontractible-loop states that are linearly independent with each other and also with the CLS. Thus, real-space topology implies there are 
\begin{equation}
    N_\text{FBS}=(6N-1) + (2N+2) =8N+1\label{eqn:N-FBS-8kag}
\end{equation} 
linearly independent states with energy $E=-2t$ in the spectrum. Comparing with Eq.~(\ref{FBtotal}) and recalling that $N_\textrm{S}=24N$, we conclude that $f=1/3$ and $w=1$ for the octagon-kagome lattice.
The presence of a nonzero band-touching index $w$ suggests that there is no energy gap separating the flat band at $E=-2t$ from the rest of the spectrum, which is indeed what we observe from numerical exact diagonalization on PBC clusters [Fig.~\ref{fig:8kagome-DoS}(a)].

Before turning to the momentum-space analysis, we note that the above real-space topological argument has also recently appeared, on a less formal level, in the master's thesis of D.~Urwyler~\cite{Urwyler:2021}. 
Furthermore, since the octagon-kagome lattice corresponds to the line graph of the bipartite $\{8,3\}$ lattice, the equality in Eq.~(\ref{eqn:N-FBS-8kag}) can also be obtained from algebraic graph-theoretic arguments discussed in Ref.~\onlinecite{saa2021}. However, we are not aware of a graph-theoretic derivation of this equality for line graphs of non-bipartite lattices or for dice lattices, making that approach inapplicable to the hyperbolic lattices analyzed in subsequent sections. Additionally, the real-space topological arguments provide a valuable physical intuition for the otherwise abstract mathematical arguments of Ref.~\onlinecite{saa2021}.

\subsection{Flat bands from hyperbolic band theory}

Having obtained the flat-band fraction $f$ and band-touching index $w$ from a purely real-space argument, we now show that those same quantities can be obtained from HBT. Furthermore, we find that the nonzero value of $w$ indeed signals a band-touching phenomenon, but in the 4D momentum space of HBT.

\subsubsection{Abelian states at zero momentum: band touching}\label{sec:8kag-HBT-1}

The $24\times 24$ Bloch Hamiltonian $H(\b{k})$ that gives the Abelian HBT spectrum of the octagon-kagome lattice can be constructed using the methods of Ref.~\cite{Boettcher:2021}: for each bond that crosses the $\gamma_j$ ($\gamma_j^{-1}$) boundary segment of the Bolza cell in Fig.~\ref{fig:8kagome-sites}, we multiply the hopping amplitude $t$ by a Bloch phase factor $e^{ik_j}$ ($e^{-ik_j}$). Focusing first on the spectrum at zero crystal momentum, $\b{k}=(k_1,k_2,k_3,k_4)=(0,0,0,0)$, the Bloch Hamiltonian $H(\b{0})$ reduces to the exact hopping Hamiltonian on a single Bolza cell with PBC~\cite{Maciejko:2021}. Out of the 24 resulting eigenenergies, we find that 9 are at the flat-band energy $E=-2t$. This is expected from the real-space argument above, summarized by Eq.~(\ref{eqn:N-FBS-8kag}), as applied to a PBC cluster with $N=1$ Bolza cells.

The 9 linearly independent Bloch states with $\b{k}=\b{0}$ can be explicitly constructed as follows. The 6 CLS of Fig.~\ref{fig:8kagome-CLS} restricted to the central Bolza cell directly give 6 such Bloch states, 5 of which are linearly independent. Additionally, noncontractible-loop states can be constructed as exemplified along the green path in Fig.~\ref{fig:8kagome-homology}: we note that the two blue sites on the $\gamma_2,\gamma_2^{-1}$ boundaries of the Bolza cell have the same amplitude $-1$, as required by the $\b{k}=\b{0}$ boundary conditions. There is one such state for each side-pairing generator $\gamma_1,\ldots,\gamma_4$, for a total of 4. We verify that all $5+4=9$ Bloch states so constructed with $\b{k}=\b{0}$ have energy $E=-2t$ and are linearly independent.

\begin{figure}[t!]
\centering
    \includegraphics[width=0.765\linewidth]{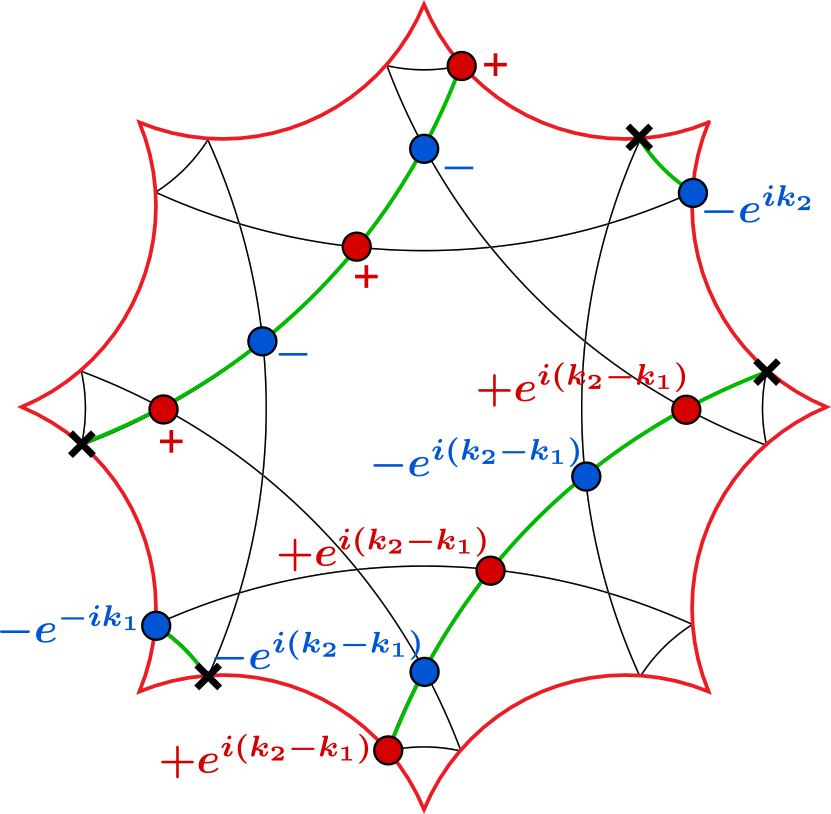}
    \caption[]{Example flat-band state that winds nontrivially around a single Bolza cell with TBC  
    captured by 4-momentum $\bs{k}=(k_1,k_2,k_3,k_4)$ (cf.~Fig.~\ref{fig:8kagome-sites} for the correspondence between momenta and the generators of Fuchsian translations). The green path is a geodesic on the Bolza surface that crosses each boundary an even number of times.
    }
\label{fig:8kagome-homology-TBC}
\end{figure}

\subsubsection{Abelian states at nonzero momentum}\label{sec:8kagome_k!=0}

\begin{figure}[t!]
\centering
    \includegraphics[width=0.99\linewidth]{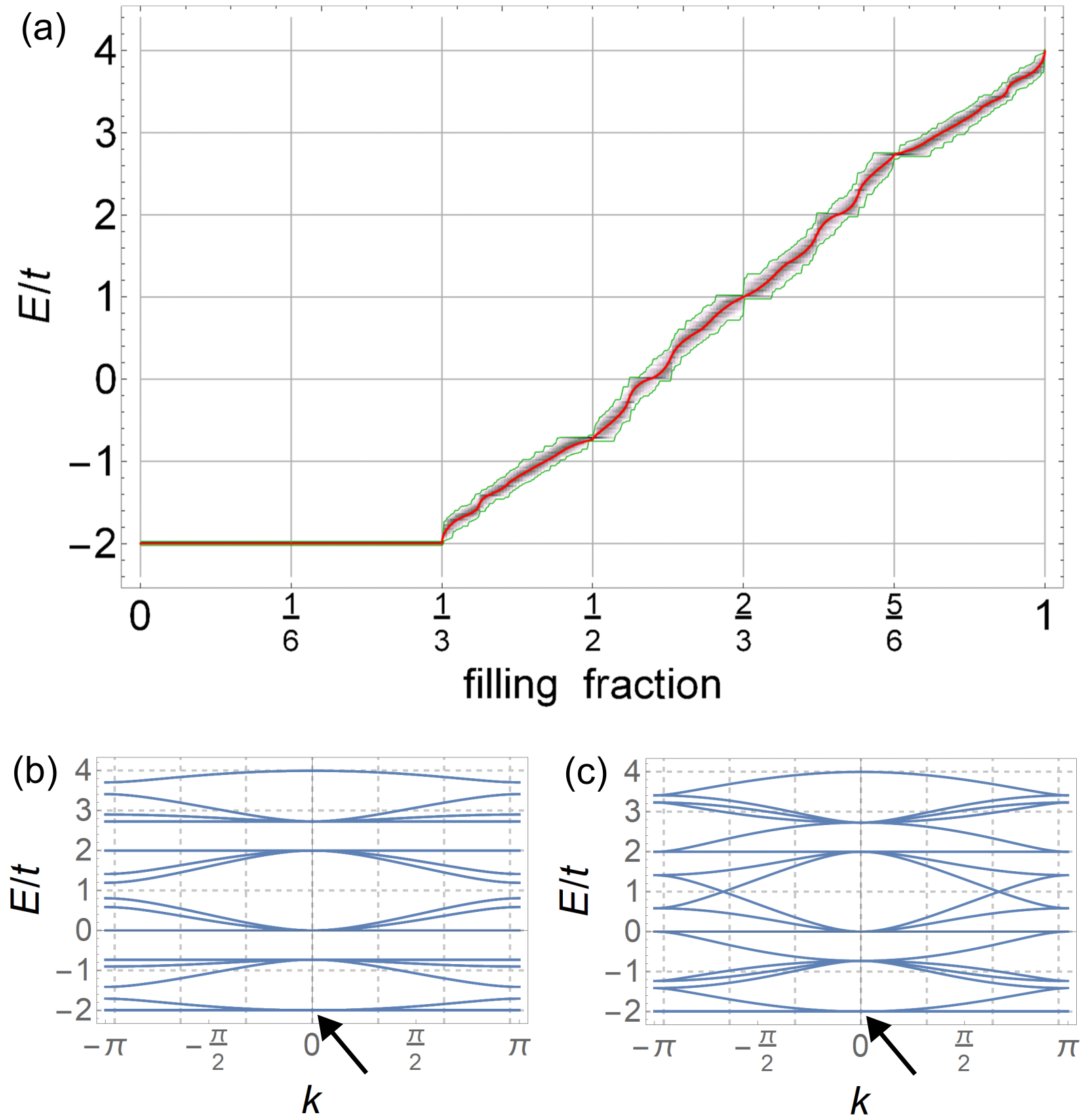}
    \caption[]{
    Spectrum of the octagon-kagome lattice.
    (a)~HBT spectrum (red) from $10^4$ momentum points in the 4D Brillouin zone vs.~exact spectrum (grayscale density plot bounded by the green range) from 8,544 PBC clusters with $N=24$ Bolza cells (576 sites), of which 1,560 are non-Abelian.
    (b,c)~Abelian band dispersion along high-symmetry lines $\bs{k}=k\bs{n}$ in the 4D Brillouin zone. The chosen directions are $\bs{n}\in\{(1,0,0,0),(1,1,0,0),(1,1,1,0)\}$ in~(b), and $\bs{n}\in\{(1,0,1,0),(1,1,1,1)\}$ in~(c). Arrows indicate the touching of the flat bands at $E=-2t$ with a dispersive band at $\bs{k}=\bs{0}$.
    }
\label{fig:8kagome-DoS}
\end{figure}

Diagonalizing the Bloch Hamiltonian $H(\b{k})$ for generic $\b{k}$, we find 8 flat bands with $E(\b{k})=-2t$ in 4D momentum space, that touch a single dispersive band at $\b{k}=\b{0}$. Since there are 24 Abelian bands in total, and $N$ independent Bloch states per band for a system with $N$ unit cells, we obtain Eq.~(\ref{FBfrac1}) with $f=1/3$ and $w=1$, as in the real-space analysis. This validates our interpretation of the parameter $w$ as an index characterizing band touching in the 4D momentum space of Abelian HBT. 
To visualize the touching, we plot in Fig.~\ref{fig:8kagome-DoS}(b,c) the Abelian band structure of the octagon-kagome lattice along those high-symmetry lines of the 4D Brillouin zone which pass through $\bs{k}=\bs{0}$.

In analogy with the $\b{k}=\b{0}$ case, we can explicitly construct the $\b{k}\neq\b{0}$ Bloch states that span the flat band by considering a single Bolza cell with $\b{k}$-dependent, ``twisted'' boundary conditions (TBC). Six Bloch states correspond to the 6 CLS of Fig.~\ref{fig:8kagome-CLS}, but the site amplitudes are now muliplied by $\b{k}$-dependent phase factors to account for the TBC. Because of these nontrivial phase factors, the 6 single-octagon states are now linearly independent. On the other hand, the noncontractible-loop states that wind across the Bolza surface once, defined along the green path in Fig.~\ref{fig:8kagome-homology}, are now incompatible with the TBC. However, we can construct 4 linearly independent states that cross each boundary of the Bolza cell an even number of times. One such state is explicitly displayed in Fig.~\ref{fig:8kagome-homology-TBC}, while the other three correspond to analogous states rotated by $2\pi/8$ around the center of the Bolza cell (with properly adjusted momentum phase factors). 
Out of the $6+4=10$ Bloch states so constructed, we find that only 8 are linearly independent for nonzero $\b{k}$, and thus span the flat band.

\subsubsection{Non-Abelian states}
\label{sec:8kag_NA_FBS}

Having established both Eq.~(\ref{FBtotal}) and Eq.~(\ref{FBfrac1}) from the real-space and 4D momentum-space perspectives, respectively, we can subtract them to obtain Eq.~(\ref{FBfrac2}). This latter equation means that the non-Abelian states do not participate in the band touching, and that they are characterized by the same flat-band fraction $f$ as the Abelian states.

As opposed to the Abelian case, where the space of 1D irreps admits a simple parametrization in terms of a 4D crystal momentum, there is no known explicit parametrization of the space of higher-dimensional irreps (although if an irrep is explicitly known, non-Abelian Bloch eigenstates can be constructed for it~\cite{cheng2022}). In the absence of an explicit parametrization of the non-Abelian Brillouin zones, we rely on numerical exact diagonalization of PBC clusters. A PBC cluster of $N$ sites is termed Abelian or non-Abelian according to whether the factor group $\Gamma_{\{8,8\}}/\Gamma_\text{PBC}$ of order $N$ is Abelian or non-Abelian. Since Bloch states on a PBC cluster are classified by irreps of this group, the spectrum of Abelian PBC clusters consists entirely of Abelian states, while non-Abelian PBC clusters possess both Abelian and non-Abelian states. (In particular, all PBC clusters with a prime number $N$ of unit cells are Abelian, since $\Gamma_{\{8,8\}}/\Gamma_\text{PBC}$ is then isomorphic to the Abelian group $\mathbb{Z}_N$~\cite{Maciejko:2022}.)

We consider PBC clusters with up to $N=24$ 
Bolza cells (576 sites). 
To mitigate the computational cost, we reduce the number of PBC clusters considered by selecting only those that form a connected region before compactification, and for which the transversal for $\Gamma_\text{PBC}$ in $\Gamma_{\{8,8\}}$ only contains words of length up to two~\cite{Maciejko:2022}. 
For each cluster, we compute the exact spectrum. Additionally for non-Abelian clusters, we separately compute the total number $N^\text{ab}$ of Abelian states in the spectrum, and the number $N^\text{ab}_\text{FBS}$ of Abelian states in the flat band at $E=-2t$. To perform this analysis~\cite{Maciejko:2022}, we first compute in GAP the character table of the finite group $\Gamma_{\{8,8\}}/\Gamma_\text{PBC}$, and extract from it the number $\c{N}_\text{1D}$ and characters $\chi^{(\lambda)}(g)$, $\lambda=1,\ldots,\c{N}_\text{1D}$ of 1D irreps of $\Gamma_{\{8,8\}}/\Gamma_\text{PBC}\ni g$. Using the characters and representation matrices for $\Gamma_{\{8,8\}}/\Gamma_\text{PBC}$, we then construct projector matrices $\Pi^{(\lambda)}$ that commute with the hopping Hamiltonian and allow us to block diagonalize it in sectors belonging to distinct irreps.

In Fig.~\ref{fig:8kagome-DoS}(a), we plot the exact spectrum for all 8,544 PBC clusters constructed with $N=24$, together with the HBT spectrum of Abelian states obtained from random sampling of $\b{k}$ points in the 4D Brillouin zone. Besides the fact that the Abelian HBT spectrum captures the overall features of the exact spectrum remarkably well, we find indeed that the flat-band fraction of non-Abelian states is again 1/3, for all 1,560 non-Abelian clusters at $N=24$. This is true despite the fact that the relative proportion of Abelian to non-Abelian states varies from cluster to cluster, and holds for all values of $N\in\{12,16,18,20,21,24\}\leq 24$ that support non-Abelian clusters.

\section{Octagon-dice lattice}
\label{sec:8dice}

\begin{figure}[t!]
\centering
    \includegraphics[width=0.855\linewidth]{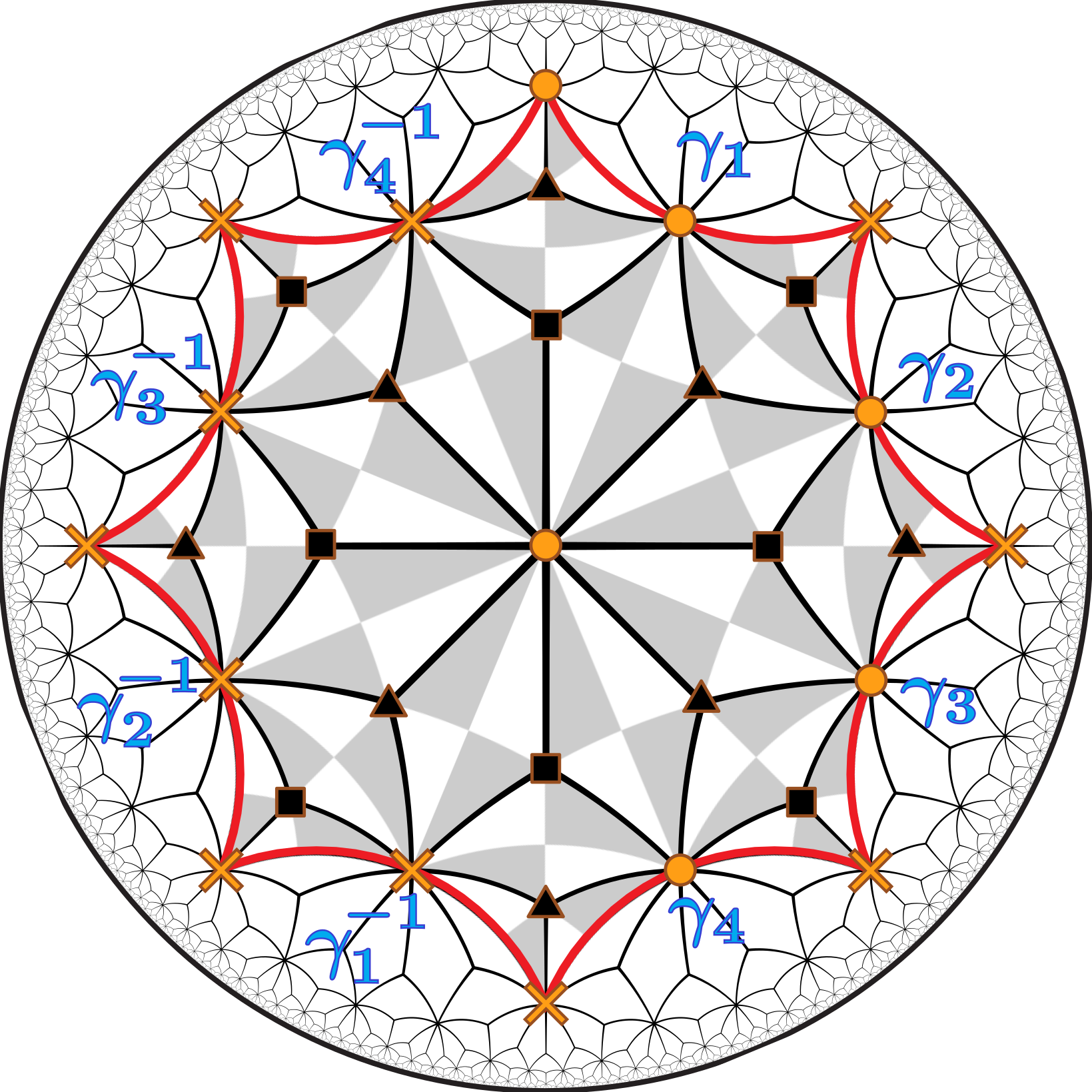}
    \caption[]{
    Octagon-dice lattice. The Bolza cell and Fuchsian group generators are as in Fig.~\ref{fig:8kagome-sites}, but the cell now contains 22 inequivalent sites. The lattice is bipartite, with 16 three-coordinated sites (black symbols) and 6 eight-coordinated sites (orange dots) per Bolza cell.
    Crossed sites on the boundary of the displayed cell belong to adjacent Bolza cells.
    Note that the three-coordinated sites constitute the $\{8,3\}$ lattice (see also Fig.~\ref{fig:8dice_CLS}), which is itself bipartite as indicated by the black square vs.~triangle symbols.
    }
\label{fig:8dice-sites}
\end{figure}

Having seen how the flat-band fraction $f=1/3$ and band-touching index $w=1$ in the octagon-kagome lattice arise from the interplay of real-space topology and Abelian/non-Abelian Bloch theory in hyperbolic space, we now turn to another hyperbolic lattice with octagonal symmetry: the octagon-dice lattice (Fig.~\ref{fig:8dice-sites}). 
This lattice can be viewed as a hyperbolic analog of the Euclidean, 6-fold-symmetric dice lattice~\cite{Sutherland:1986}. 
Together with the heptagon-dice lattice to be studied last (Sec.~\ref{sec:7dice}), these dice (also called ``rhombille'') lattices illustrate a mechanism for flat-band physics that is distinct from the line-graph mechanism of kagome lattices: chiral flat bands on bipartite lattices with sublattice (chiral) asymmetry~\cite{Sutherland:1986,lieb1989,ramachandran2017}.

To understand this mechanism, we first observe that the octagon-dice lattice is bipartite, with hopping only between three-coordinated sites and eight-coordinated sites (black vs.~orange symbols in Fig.~\ref{fig:8dice-sites}, respectively). 
Each Bolza cell contains 16 three-coordinated sites ($A$ sublattice) and 6 eight-coordinated sites ($B$ sublattice), for a total of 22 sites per cell. A PBC cluster with $N$ cells thus contains $N_A=16N$ three-coordinated sites and $N_B=6N$ eight-coordinate sites, for a total of $N_A+N_B=22N$ sites. The hopping Hamiltonian $\c{H}$ can be written in block form as:
\begin{align}\label{chiralH}
    \c{H}=\left(\begin{array}{cc}
    \mathbbold{0} & M \\
    M^\dag & V_B\mathbbold{1}\end{array}\right),
\end{align}
where $\mathbbold{0}$ denotes the $N_A\times N_A$ zero matrix, $M$ a $N_A\times N_B$ matrix which contains the hopping amplitudes, and $M^\dag$ its $N_B\times N_A$ Hermitian conjugate. We have also added an on-site potential of strength $V_B$ on the $B$ sublattice only ($\mathbbold{1}$ stands here for the $N_B\times N_B$ identity matrix). 

The appearance of the flat-band states can now be understood through the rank-nullity theorem as follows. 
Denote by $\tilde{\psi}_A\in\mathrm{null}(M^\dag)$ a right zero eigenvector of $M^\dag$, and by $\tilde{\psi}_B\in\mathrm{null}(M)$ a right zero eigenvector of $M$. 
Likewise, denote the dimension of those null spaces by $\tilde{N}_A=\dim\mathrm{null}(M^\dag)$ and $\tilde{N}_B=\dim\mathrm{null}(M)$. The rank-nullity theorem implies that
\begin{equation}
N_A = \mathrm{rank}(M^\dag) + \tilde{N}_A \quad\textrm{and} \quad N_B = \mathrm{rank}(M) + \tilde{N}_B.   
\end{equation}
However, since the rank of the rectangular matrices $M$ and $M^\dagger$ must match, we have that $N_A-\tilde{N}_A=N_B-\tilde{N}_B$, from which we derive the inequality $\tilde{N}_A\geq N_A-N_B$. Now, for any $\tilde{\psi}_A\in\mathrm{null}(M^\dag)$, the vector:
\begin{align}\label{PsiA}
    \tilde{\Psi}_A=\left(\begin{array}{c}
    \tilde{\psi}_A \\
    0
    \end{array}\right),
\end{align}
is a zero eigenvector of $\c{H}$ with support on the $A$ sublattice only. The Hamiltonian (\ref{chiralH}) thus has a flat band at $E=0$ with degeneracy $N_\text{FBS}$ that obeys a lower bound given by the sublattice imbalance~\cite{Sutherland:1986,lieb1989}:
\begin{align}\label{chiralFBS}
    N_\text{FBS}\geq N_A-N_B,
\end{align}
in the convention where $A$ is the majority sublattice. For the octagon-dice lattice, Eq.~(\ref{chiralFBS}) implies $N_\text{FBS}\geq 10N$. Since $N_\text{S}=N_A+N_B=22N$, this result implies the lower bound $f\geq 5/11$ (as well as the trivial bound $w\geq 0$). To determine the precise values of $f$ and $w$, we turn to our analysis based on real-space topology and HBT.

\subsection{Flat bands from real-space topology}
\label{sec:8dice-CLS}

\begin{figure}[t!]
\centering
    \includegraphics[width=0.855\linewidth]{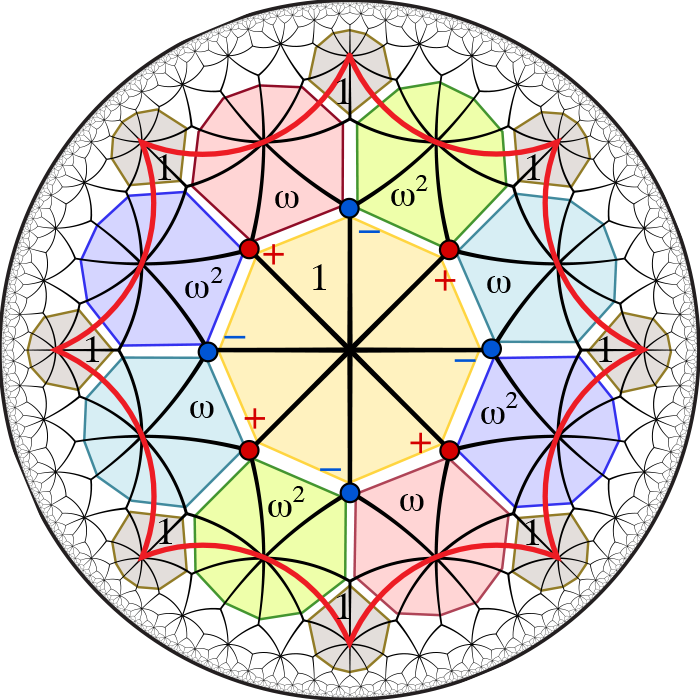}
    \caption[]{Maximally localized CLS on the octagon-dice lattice. Octagons shaded in six distinct colors represent the six distinct CLS belonging to one Bolza cell, with nonvanishing amplitudes $\pm 1$ (red/blue dots) on 3-coordinated sites only. 
    For each CLS we define the amplitude to be positive (negative) on the sites indicated with a triangle (square) in Fig.~\ref{fig:8dice-sites}.
    A superposition of all CLS with coefficients $1,\omega,\omega^2$ as indicated produces zero, where $\omega=e^{\pm i2\pi/3}$.}
\label{fig:8dice_CLS}
\end{figure}

We first consider CLS associated to contractible loops on the lattice. As Eq.~(\ref{PsiA}) suggests, we look for CLS with support on 3-coordinated sites. As advertised earlier, the maximally localized CLS is again a single-octagon state, bound to a single face of the $\{8,3\}$ lattice (Fig.~\ref{fig:8dice_CLS}). The probability amplitude alternates between $+1$ and $-1$ around the 8 sites surrounding an 8-coordinated vertex, generalizing the single-hexagon CLS of the Euclidean dice lattice~\cite{Sutherland:1986}. As before, there are 6 inequivalent single-octagon states per Bolza cell, leading to a total of $6N$ zero-energy states for a PBC cluster with $N$ sites. 
However, there are now 2 independent constraints among those states, corresponding to linear superpositions of all states with $\omega$-dependent coefficients as illustrated in Fig.~\ref{fig:8dice_CLS}, where $\omega=e^{\pm i2\pi/3}$ is one of the two nontrivial cube roots of unity. Since each 3-coordinated site is shared by 3 single-octagon states in this superposition, the distribution of coefficients ensures that the overall amplitude vanishes at each such site, since $1+\omega+\omega^2=0$. As a result, there are $6N-2$ linearly independent CLS on an $N$-cell PBC cluster.

\begin{figure}[t!]
\centering
    \includegraphics[width=0.755\linewidth]{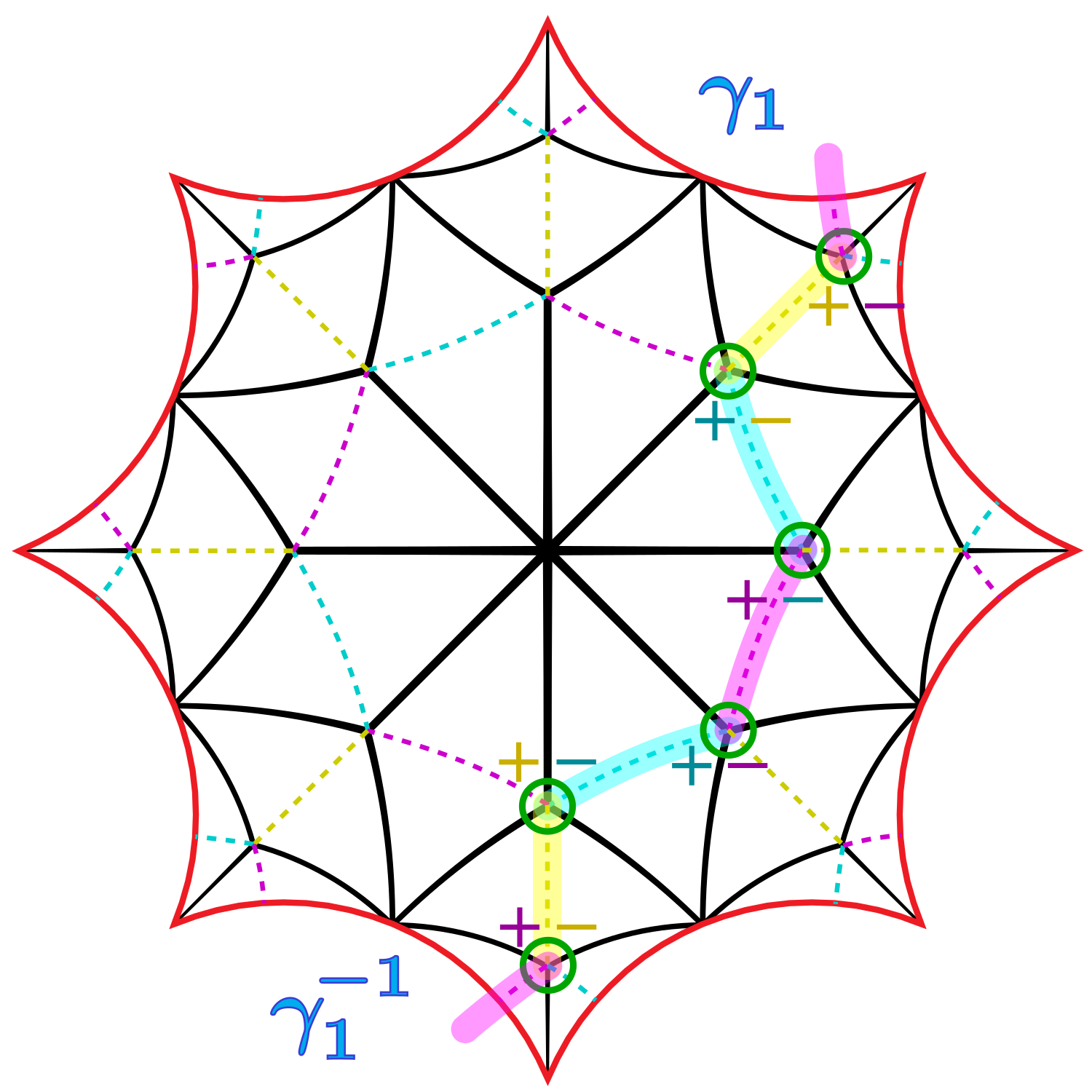}
    \caption[]{Three flat-band states (indicated by shades of cyan/magenta/yellow) that wind around the Bolza cell along a noncontractible loop corresponding to the $\gamma_1$ direction. 
    Each state has nonvanishing amplitude on exactly four of the six sites (green circles) along the highlighted string. 
    The sum of all three states vanishes; therefore, only two of them are linearly independent. 
    Analogous noncontractible-loop states can be constructed for nontrivial cycles corresponding to $\gamma_{2,3,4}$. 
    }
\label{fig:8dice_homology-colored}
\end{figure}

Next, we consider noncontractible-loop states. Three examples of such states for $N=1$ (single Bolza cell) are shown in Fig.~\ref{fig:8dice_homology-colored}.
These states are constructed according to the following prescription.
First, augment the dice lattice with the missing short diagonals of the rhombi (dashed lines in the figures) and color them cyan, magenta, or yellow such that:
(1)~lines of all three colors meet at each 3-coordinated site, and 
(2)~each 8-coordinated site is enclosed in a dashed octagon made of two colors only. 
We explicitly show in Fig.~\ref{fig:8dice_homology-colored} one such coloring for a single Bolza cell which is also compatible with the Fuchsian translation group; consequently, by applying Fuchsian translations one generates a coloring of the diagonal segments compatible with the rules (1) and (2) also for an $N$-cell PBC cluster.
Second, select a noncontractible loop passing along the cyan, magenta, and yellow segments (but not along the physical bonds displayed in black).
Finally, to construct a noncontractible-loop state, choose one of the three colors and a direction along the loop, and place amplitude $+1$ ($-1$) at the beginning (at the end) of each diagonal segment of the selected color.
For a given oriented loop, this procedure produces three different states (whose amplitudes in Fig.~\ref{fig:8dice_homology-colored} are indicated by the dark cyan/magenta/yellow $\pm$ symbols). However, one easily verifies that only two of these states are linearly independent. Indeed, as each site along the loop is simultaneously at the beginning of one and at the end of another diagonal segment, it follows that the sum of the three states results in the zero vector. 

The above construction can be repeated for each inequivalent noncontractible loop. 
As determined below Eq.~(\ref{eqn:Bolza-cluster-genus}) in the context of the octagon-kagome lattice, we know that an $N$-cell PBC cluster is compactified on a genus-$(N+1)$ surface that supports $2h=2(N+1)$ inequivalent noncontractible loops. Given that there are two independent noncontractible-loop states per each such loop, real-space topology predicts a total of 
\begin{equation}
N_\textrm{FBS} = (6N-2) + 4(N+1) = 10 N +2    
\end{equation}
linearly independent flat-band states. 
Using $N_\text{S}=22N$ and comparing with Eq.~(\ref{FBtotal}), we find a flat-band fraction $f=10/22=5/11$, which saturates the lower bound found earlier, and a band-touching index of $w=2$. As for the octagon-kagome lattice, since $w\neq 0$, we expect that the flat band is not gapped, which is consistent with the exact spectrum obtained numerically [Fig.~\ref{fig:8dice-DoS}(a)].

\subsection{Flat bands from hyperbolic band theory}

As for the octagon-kagome lattice, to further elucidate the meaning of the extra $w=2$ states, we analyze the problem from the point of view of HBT.

\subsubsection{Abelian states at zero momentum: band touching}

The Bloch Hamiltonian $H(\b{k})$ that captures Abelian states is now a $22\times 22$ matrix. If the potential $V_B$ on the $B$ sublattice is set to zero, the Bloch Hamiltonian $H(\b{0})$ has 14 zero eigenvalues. However, for arbitrary $V_B\neq 0$, the degeneracy is partially lifted to 12. Since a nonzero $V_B$ preserves the symmetries of the lattice, this is the generic case we consider henceforth. (Note that an on-site potential on the $A$ sublattice, $V_A$, can always be regarded as a shifted $B$ sublattice potential $\tilde{V}_B=V_B-V_A$ together with a rigid shift of all bands by an amount $V_A$, which preserves the band flatness.)

Twelve linearly independent Bloch states with $\b{k}=\b{0}$ can be constructed as follows. The 6 single-octagon CLS of Fig.~\ref{fig:8dice_CLS} are consistent with PBC on the Bolza cell, but only 4 are linearly independent because of the 2 constraints discussed earlier. We further find eight additional linearly independent flat-band states that wind nontrivially around the Bolza cell, namely two per each of the four nontrivial loops (corresponding to translations by $\gamma_{1,2,3,4}$). 
These are precisely the noncontractible-loop states described in the previous section (Fig.~\ref{fig:8dice_homology-colored}). 
We verify that the resulting $2\times 4=8$ states are linearly independent of each other as well as from the 4 linearly independent CLS in Fig.~\ref{fig:8dice_CLS}. 
These $4+8=12$ states thus correctly account for the 12-fold degeneracy of the Bloch Hamiltonian at $\b{k}=\b{0}$ for arbitrary $V_B\neq 0$.

\begin{figure}[t!]
\centering
    \includegraphics[width=0.853\linewidth]{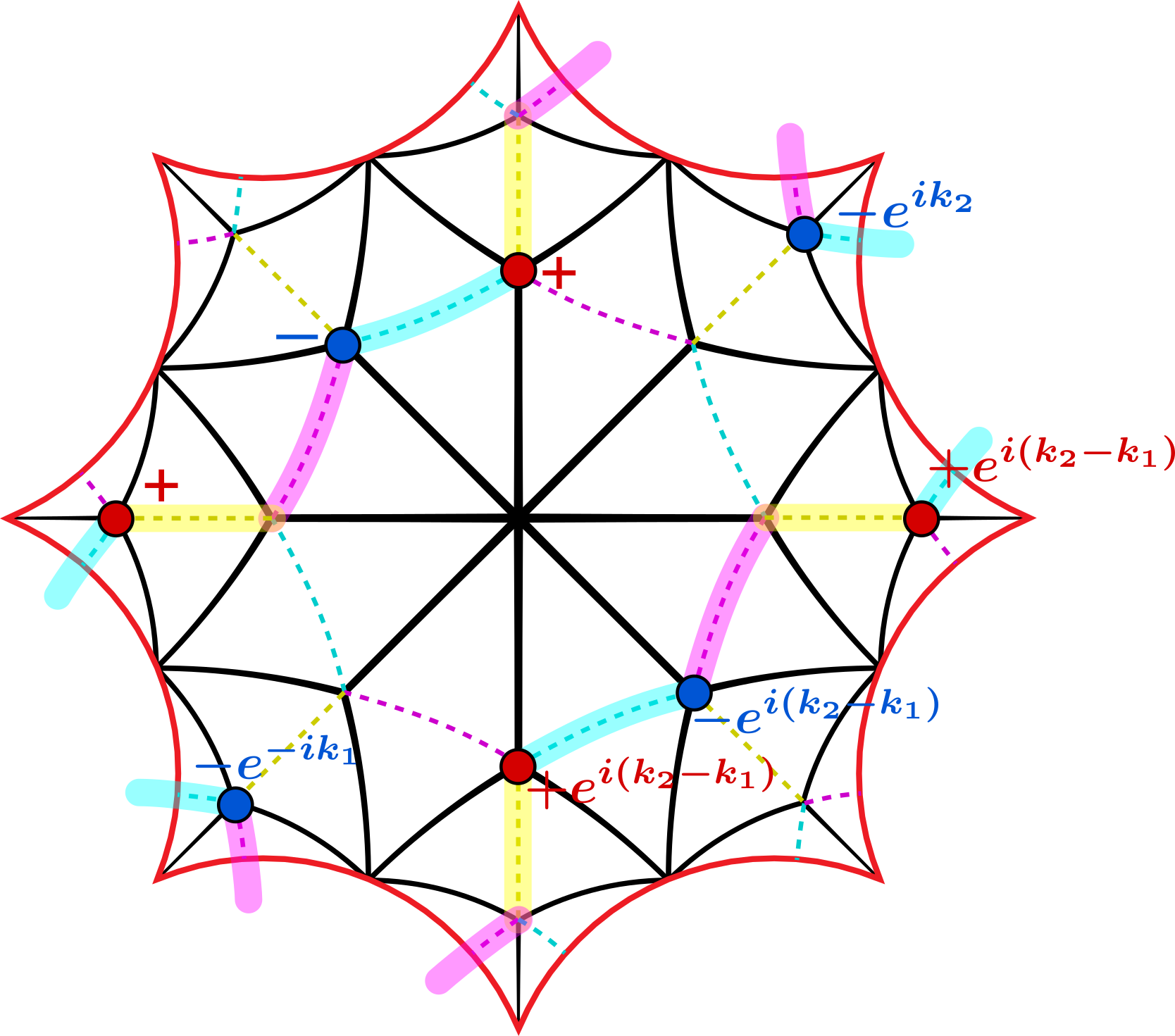}
    \caption[]{Example flat-band state on the octagon-dice lattice that winds nontrivially around a single Bolza cell with TBC  
    captured by 4-momentum $\bs{k}=(k_1,k_2,k_3,k_4)$, see~Fig.~\ref{fig:8dice-sites}. 
    }
\label{fig:8dice-homology-TBC}
\end{figure}

\subsubsection{Abelian states at nonzero momentum}

Out of the 22 bands in the 4D Brillouin zone, we find 10 flat bands with $E(\b{k})=0$, regardless of the value of the $B$ sublattice potential $V_B$. 
We further find that for $V_B=0$ the flat bands touch 4 dispersive bands at $\bs{k}=\bs{0}$; however, for the generic case with $V_B\neq 0$ we find that 2 out of these 4 bands detach from the touching point. 
Therefore, we conclude that the touching index $w=2$, as determined in Sec.~\ref{sec:8dice-CLS} from the real-space topology, is also reproduced by the HBT calculations in the 4D Brillouin zone.
To visualize the touching, we plot in Fig.~\ref{fig:8dice-DoS}(b,c) the Abelian band structure of the octagon-dice lattice along high-symmetry lines that pass through point $\bs{k}=\bs{0}$ of the 4D Brillouin zone.

To explicitly construct the flat-band Bloch states with $\bs{k}\neq\bs{0}$, we again dress the real-space CLS which cross the boundary of the Bolza cell with $\bs{k}$-dependent phase factors to satisfy the TBC of Abelian HBT.
This produces 6 single-octagon states which are linearly independent at generic $\bs{k}$, in analogy with our finding for the octagon-kagome lattice in Sec.~\ref{sec:8kagome_k!=0}.
Additionally, we consider the two noncontractible-loop states along the loop, shown in Fig.~\ref{fig:8dice-homology-TBC}, which crosses each boundary of the Bolza cell an even number of times.
We finally consider three additional copies of both those noncontractible-loop states, which are obtained by rotating the trajectory in Fig.~\ref{fig:8dice-homology-TBC} by $2\pi n/8$ with $n\in\{1,2,3\}$. This results in a total of $6$ CLS and $8$ noncontractible-loop states.
We compute the rank of the $\bs{k}$-dependent matrix formed by the $6+8=14$ corresponding column eigenvectors. 
For a random sampling of $10^4$ $\b{k}$-points uniformly distributed throughout the 4D Brillouin zone, we find that the rank is always 10, confirming that these states form an (overcomplete) Bloch basis for the flat band.

\subsubsection{Non-Abelian states}

\begin{figure}[t!]
\centering
    \includegraphics[width=0.99\linewidth]{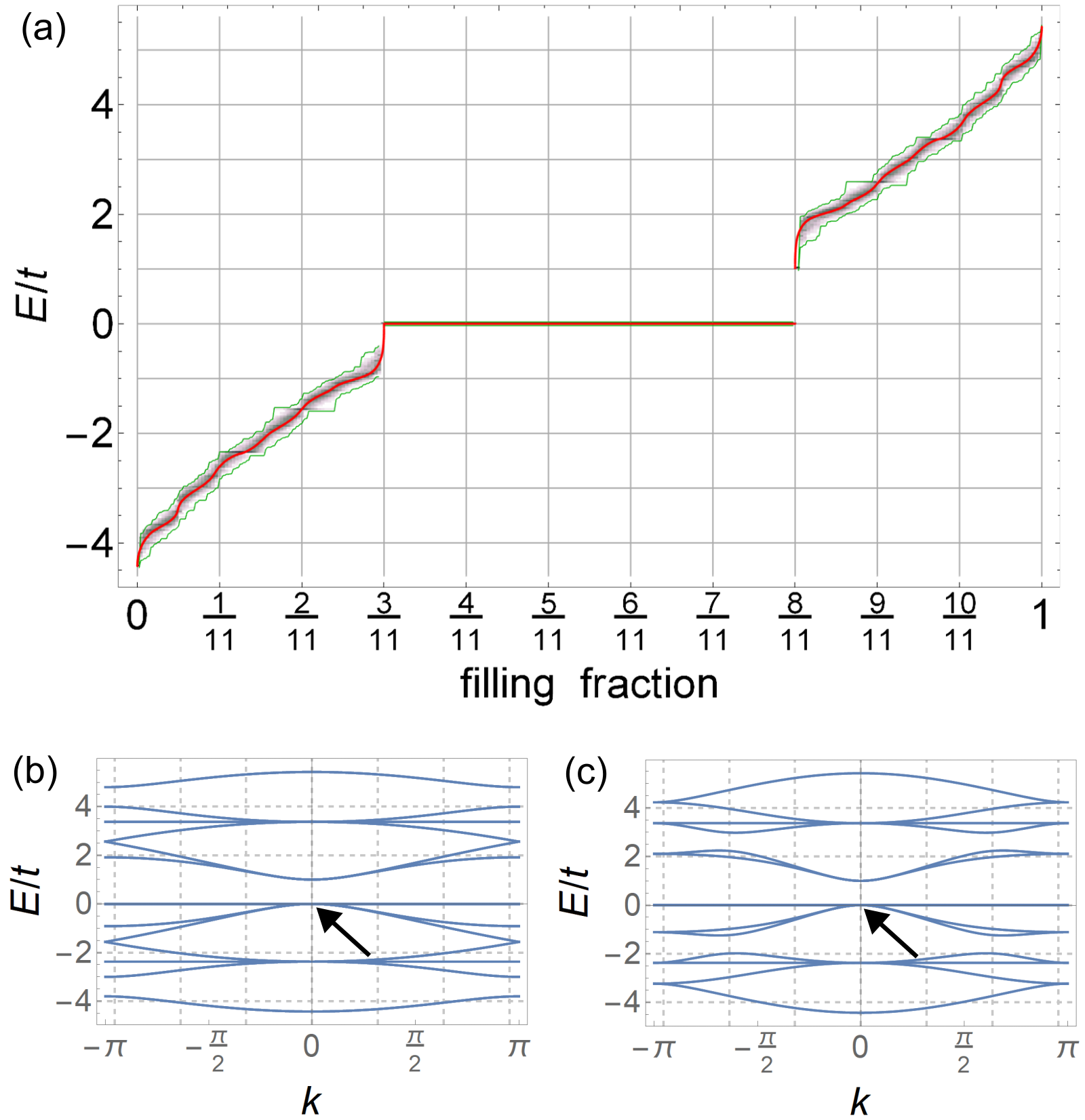}
    \caption[]{
    Spectrum of the octagon-dice lattice ($V_B/t=1$).
    (a)~HBT spectrum (red) from $2\times 10^4$ momentum points in the 4D Brillouin zone vs.~exact spectrum (grayscale density plot with support bounded by the green range) from 8,544 PBC clusters with $N=24$ Bolza cells (528 sites), of which 1,560 are non-Abelian.
    (b,c)~Abelian band dispersion along high-symmetry lines $\bs{k}=k\bs{n}$ in the 4D Brillouin zone. 
    The chosen directions are $\bs{n}\in\{(1,0,0,0),(1,1,0,0),(1,1,1,0)\}$ in~(b), and $\bs{n}\in\{(1,0,1,0),(1,1,1,1)\}$ in~(c). Arrows indicate the touching of the flat bands at $E=0$ with two dispersive bands at $\bs{k}=\bs{0}$.
    }
\label{fig:8dice-DoS}
\end{figure}

In Fig.~\ref{fig:8dice-DoS}(a), we plot aggregate data for the spectrum of the 8,544 PBC clusters of the octagon-dice lattice with $N=24$ Bolza cells, which contain 528 sites in total. We set the $B$ sublattice potential to $V_B/t=1$, which gaps out the flat band on the $E>0$ part of the spectrum but leaves the $E<0$ band touching unaffected (likewise, $V_B<0$ would gap out the $E<0$ spectrum but preserve the $E>0$ band touching.) As previously, Abelian HBT does not span the entire spectrum of the non-Abelian clusters but still captures its overall features. For the non-Abelian clusters, we verify using projector matrices that for all $N\leq 24$ clusters studied, Eq.~(\ref{FBfrac2}) holds with $f=5/11$. The Abelian spectrum obeys Eq.~(\ref{FBfrac1}) with that same $f$ and also $w=2$, thus we verify indeed that $f=5/11$ and $w=2$ for the octagon-dice lattice (Table~\ref{tab:summary}).

\section{Heptagon-kagome lattice}
\label{sec:7kag}

\begin{figure}[t!]
\centering
    \includegraphics[width=\linewidth]{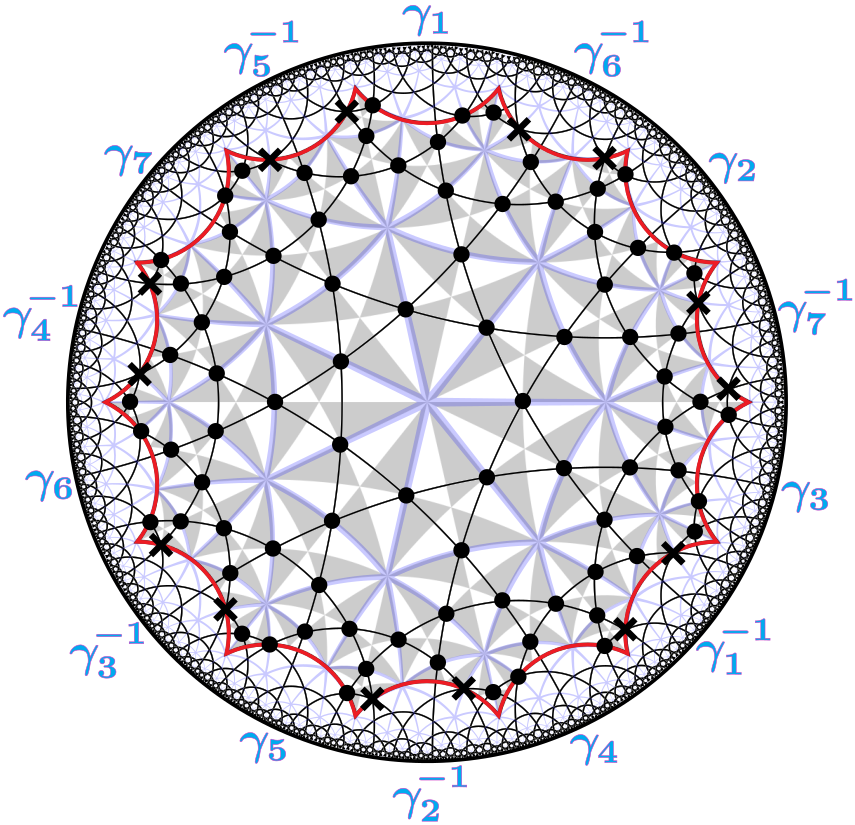}
    \caption[]{
    Heptagon-kagome lattice. The 14-sided Klein cell (red), which contains 84 inequivalent sites (black dots), can be translated by the Fuchsian group generators $\gamma_j$, $j=1,\ldots,7$ (blue) to generate the entire lattice. The 336 grey/white triangles are fundamental domains for the action of the symmetry group of the Klein surface. The edges of the underlying $\{3,7\}$ tessellation are drawn in purple.
    }
\label{fig:7kagome-sites}
\end{figure}

We now turn to hyperbolic lattices with heptagonal symmetry. In this section, we analyze the heptagon-kagome lattice (Fig.~\ref{fig:7kagome-sites}). This lattice is the line graph of the $\{7,3\}$ lattice, which is a tiling by regular heptagons, three of which meet at each vertex. As opposed to the octagonal lattices just studied, the correct translation unit cell of the $\{7,3\}$ lattice and its heptagon-kagome line graph is a regular hyperbolic 14-gon (Klein cell). In other words, the hyperbolic Bravais lattice for these two lattices is the $\{14,7\}$ lattice, whose fundamental domain is the Klein cell, 7 of which meet at each vertex~\cite{Boettcher:2021}. The corresponding Fuchsian group of translations is generated by 7 elementary generators, $\gamma_j,j=1,\ldots,7$, which identify pairs of sides of the Klein cell according to a specific pattern (Fig.~\ref{fig:7kagome-sites}). While other pairings are topologically equivalent, i.e., they all generate a genus-3 surface, this particular pairing has the advantage that it gives a Riemann surface with highly symmetric geometry, the Klein surface. This high degree of symmetry will be useful in the systematic construction of flat-band states for both the heptagon-kagome (Sec.~\ref{sec:7kagome_abelian}) and heptagon-dice (Sec.~\ref{sec:7dice_abelian}) lattices. We denote the Fuchsian group as $\Gamma_{\{14,7\}}$, and its presentation is given by~\cite{quine1996}:
\begin{align}\label{FuchsianKlein}
    \Gamma_{\{14,7\}}=\langle\gamma_1,\ldots,\gamma_7\colon&\gamma_2\gamma_4\gamma_6\gamma_1\gamma_3\gamma_5\gamma_7=\mathbbold{1},\nn\\
   & \gamma_3\gamma_6\gamma_2\gamma_5\gamma_1\gamma_4\gamma_7=\mathbbold{1}\rangle,
\end{align}
with $\mathbbold{1}$ denoting the identity element as before. The Klein cell contains 84 sites of the heptagon-kagome lattice, thus PBC clusters with $N$ Klein cells contain $84N$ sites and possess $N_\text{S}=84N$ eigenstates. Using the Riemann-Hurwitz relation, we find this time that such a PBC cluster has genus $h=N(g-1)+1=2N+1$, because the Klein surface has genus $g=3$.

\subsection{Flat bands from real-space topology}
\label{sec:7kag-CLS}

\begin{figure}[t!]
\centering
    \includegraphics[width=0.855\linewidth]{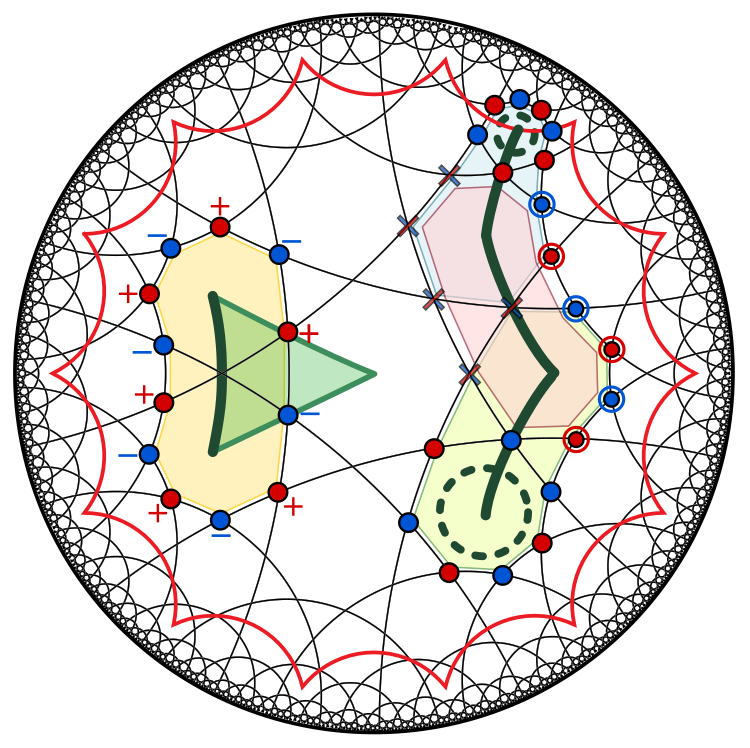}
    \caption[]{
    Left side: The maximally localized CLS on the heptagon-kagome lattice is the heptagon-pair state (left), associated to a single vertex of the heptagon-kagome lattice or, equivalently, to a single edge (dark green) of the $\{3,7\}$ lattice. 
    The equal weight superposition of the heptagon-pair CLS associated with edges of any triangle (green) of the $\{3,7\}$ lattice results in the zero vector, indicating that the constructed CLS form an overcomplete basis.
    Right side: One can construct noncontractible-loop states (dark green) as linear combinations of the heptagon-pair CLS.
    The cross symbols indicate cancellation of amplitudes from two heptagon-pair states; red/blue dots indicate amplitudes $\pm 1$; circled red/blue dots indicate amplitudes $\pm 2$.
    }
\label{fig:7kagome-CLS}
\end{figure}

As for other kagome lattices, we search for flat-band CLS at energy $E=-2t$ based on the destructive-interference principle: equal and opposite probability amplitudes $\pm 1$ on two sites of a kagome triangle give vanishing hopping amplitude onto the third. Flat-band CLS supported on contractible loops thus consist of closed chains of alternating amplitudes $\pm 1$ with a necessarily even number of sites. For the octagon-kagome lattice, the maximally localized CLS is thus supported on a single octagon of the underlying $\{8,3\}$ lattice (Fig.~\ref{fig:8kagome-CLS}). By contrast, the $\{7,3\}$ lattice that underlies the heptagon-kagome lattice is a tiling by odd-sided heptagons; thus, the maximally localized CLS must involve at least two heptagons~\cite{Kollar:2019,Kollar:2020}. We term such a CLS a {\it heptagon-pair state} (Fig.~\ref{fig:7kagome-CLS}, left side). 
Each CLS is associated to a single vertex of the  heptagon-kagome lattice or, equivalently, to a single edge of the underlying $\{3,7\}$ lattice. Therefore, there are 84 heptagon-pair CLS for each Klein cell, for a total of $84N$ CLS for a PBC cluster with $N$ unit cells.

Next, we look for constraints among those $84N$ CLS. We observe that for each triangle in the $\{3,7\}$ lattice (green triangle on left side of Fig.~\ref{fig:7kagome-CLS}), the sum of the three overlapping heptagon-pair states associated with the three edges of this triangle vanishes. The number of $\{3,7\}$ triangles in each Klein cell (which matches the number of vertices of the $\{7,3\}$ lattice in each Klein cell) is 56~\cite{Boettcher:2021}. Therefore, we have $56N$ constraints which reduce the number of linearly independent CLS to $28N$. 

Finally, we turn to flat-band states with support along noncontractible loops. In contrast with the octagon-kagome lattice, here such noncontractible-loop states can be constructed by linear superposition of an extensive number of maximally localized (heptagon-pair) CLS. 
We illustrate in Fig.~\ref{fig:7kagome-CLS} (right side) how an equal-weight superposition of three overlapping heptagon-pair states produces a CLS in which two heptagons are connected by a ``string'' or chain of alternating probability amplitudes $\pm 2$. 
Such \emph{chain CLS} appear generally in the line graphs $L(X)$ of graphs $X$ with odd-sided faces~\cite{chiu2020}. 
By adding more heptagon-pair states to the superposition, we can separate out the two heptagons arbitrarily far apart, and eventually annihilate them on a closed surface, leaving behind the string as a noncontractible-loop state. 
By contrast, due to the absence of chain CLS in the line graphs of $\{p,3\}$ lattices with $p$ even~\cite{chiu2020}, noncontractible-loop states on those line graphs (e.g., the octagon-kagome lattice of Sec.~\ref{sec:8kag}) cannot be obtained from the linear superposition of maximally localized CLS.

To summarize, real-space topology suggests that $N$-cell PBC clusters of the heptagon-kagome lattice possess $N_\text{FBS}=28N$ states in the flat band at $E=-2t$, implying a flat-band fraction $f=1/3$ as in the octagon-kagome lattice, but a vanishing band-touching index $w=0$. Thus real-space topology suggests the flat band is gapped, which is corroborated by numerical evidence [Fig.~\ref{fig:7kagome-DoS}(a)].

\subsection{Flat bands from hyperbolic band theory}

We now turn to the reciprocal-space perspective. As mentioned earlier, the dimension of momentum space in Abelian HBT depends on the genus of the compactified unit cell, which is reflected in the number of generators of the Fuchsian translation group. Although the group in Eq.~(\ref{FuchsianKlein}) has 7 generators, it obeys 2 relations, with each generator appearing once in each of the 2 relators. When performing the substitution $\gamma_j\mapsto e^{ik_j}$ in Abelian HBT, we see that both relations impose the same constraint:
\begin{align}\label{k7}
k_7=-(k_1+k_2+k_3+k_4+k_5+k_6)\mod 2\pi.
\end{align}
Therefore, the 7$^\text{th}$ component of crystal momentum is not independent, and the Brillouin zone for Abelian states is 6-dimensional (6D), spanned by the crystal momentum $\b{k}=(k_1,k_2,k_3,k_4,k_5,k_6)$.

\subsubsection{Abelian states}
\label{sec:7kagome_abelian}

\begin{figure}[t!]
\centering
    \includegraphics[width=0.99\linewidth]{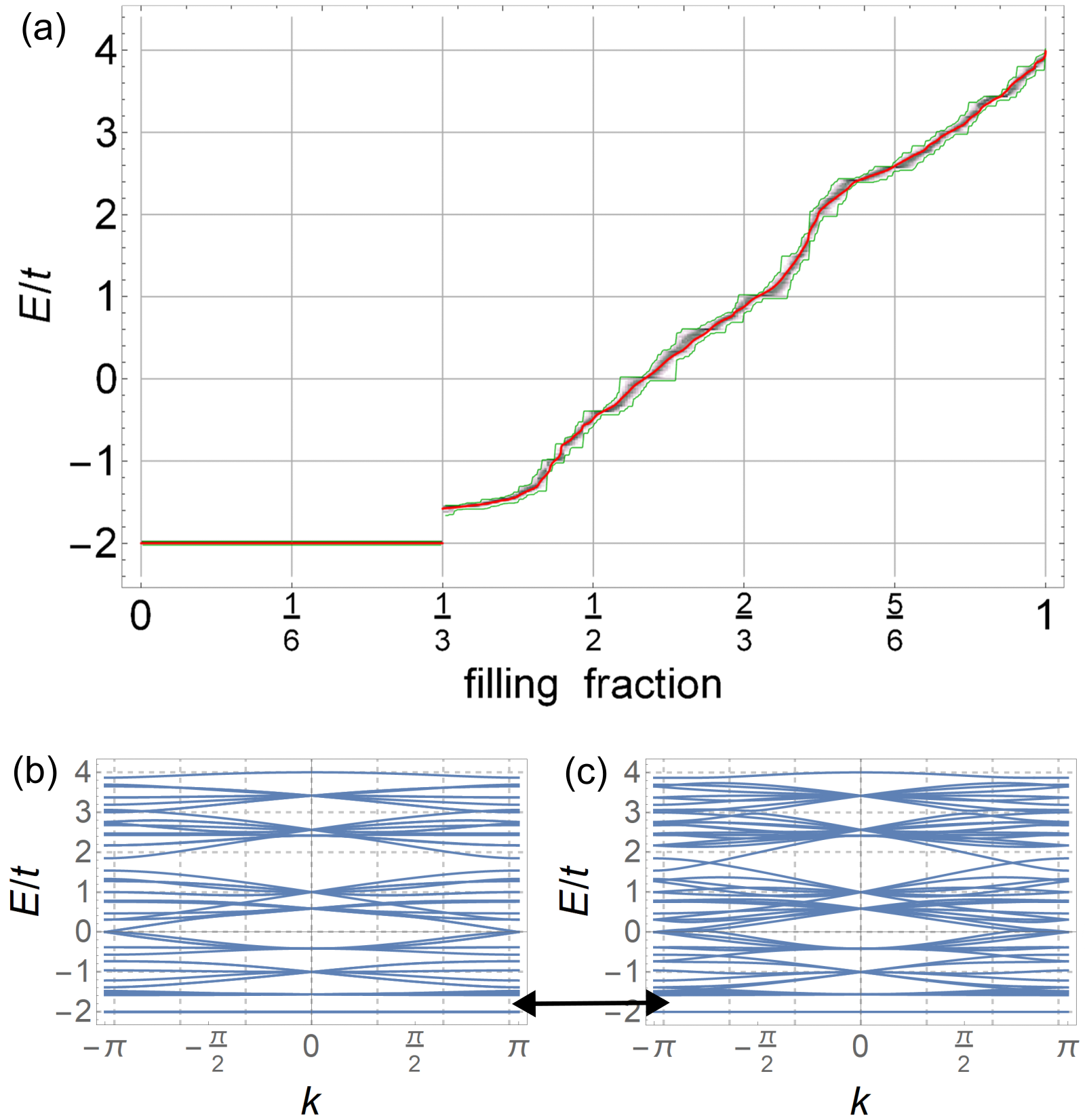}
    \caption[]{
    Spectrum of the heptagon-kagome lattice.
    (a)~HBT spectrum (red) from $10^4$ momentum points in the 6D Brillouin zone vs.~exact spectrum (grayscale density plot bounded by the green range) from 27,995 PBC clusters with $N=8$ Klein cells (672 sites), of which 2,240 are non-Abelian.
    (b,c)~Abelian band dispersion along selected high-symmetry lines $\bs{k}=k\bs{n}$ in the 6D Brillouin zone. The chosen directions are $\bs{n}=(1,0,0,0,0,0)$ in~(b), and $\bs{n}=(1,0,-1,0,1,0)$ in~(c). 
    Arrow indicates the energy gap that separates the flat bands at $E=-2t$ from the dispersive bands.
    }
\label{fig:7kagome-DoS}
\end{figure}

Diagonalization of the $84\times 84$ Bloch Hamiltonian $H(\b{k})$ reveals 28 flat bands with $E(\b{k})=-2t$ in the 6D Brillouin zone that do not touch any dispersive bands, as expected from the vanishing band-touching index $w=0$ predicted by real-space topology. 
The presence of a gap separating the flat band from the dispersive bands is visualized in Fig.~\ref{fig:7kagome-DoS}(b,c), which shows the Abelian band structure of the heptagon-kagome lattice along selected lines in the 6D Brillouin zone.
Since there are no special band-touching points with enhanced degeneracy, we directly focus on constructing flat-band hyperbolic Bloch states with generic $\b{k}$.

As previously, we wish to construct a Bloch basis for the flat bands by dressing the real-space CLS with $\b{k}$-dependent phase factors to obey TBC on a single Klein cell. The analysis is more involved than for previous lattices, as there are now 84 heptagon-pair states. To systematize the process, we exploit the high degree of symmetry of the Klein surface~\cite{klein1878,EightfoldWay}. We begin with an initial heptagon-pair state with support entirely inside the original Klein cell, e.g., the CLS illustrated on the left side of Fig.~\ref{fig:7kagome-CLS}. This CLS does not cross the boundary of the Klein cell and thus does not carry any $\b{k}$-dependent phase factors. From this initial state, we wish to construct the remaining 83 flat-band states by symmetry, which involves shifting the point of origin of the state while correctly imposing TBC on the Klein surface.

\begin{figure}[t!]
\centering
    \includegraphics[width=0.855\linewidth]{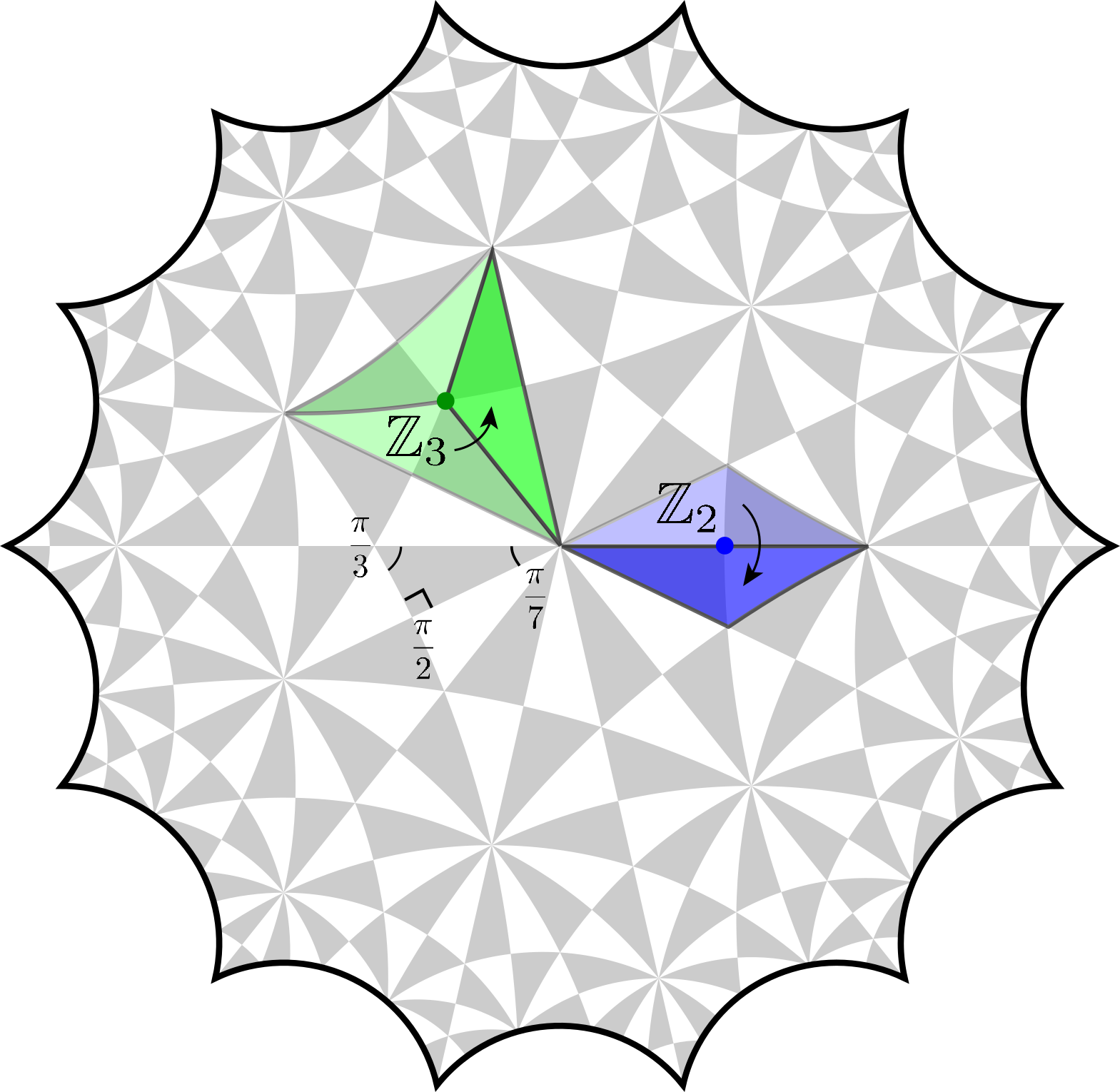}
    \caption[]{The Klein quartic is tiled by 336 Schwarz triangles (displayed in white and grey) with interior angles $\pi/2$, $\pi/3$, and $\pi/7$. 
    A pair of Schwarz triangles (highlighted blue or green triangles) is a fundamental domain for the action of the order-168 group $G\cong\mathrm{PSL}(2,\mathbb{Z}_7)$ of conformal automorphisms of the Klein quartic. Each site of the heptagon-kagome lattice (blue dot) belongs to a rhombus consisting of two such fundamental triangles related by a $\pi$ rotation around the site, and each three-coordinated site of the heptagon-dice lattice (green dot) belongs to an equilateral triangle consisting of three fundamental triangles related by a $2\pi/3$ rotation around the site.}
\label{fig:klein}
\end{figure}

To understand how to do this systematically, we first describe the symmetries of the Klein quartic, which can be interpreted as point-group symmetries of the heptagon-kagome lattice (Fig.~\ref{fig:klein}). While the entire Poincar\'e disk is tiled by infinitely many copies of the 14-sided Klein cell, the Klein cell itself is tiled by 336 copies of a hyperbolic triangle with interior angles $\pi/2$, $\pi/3$, and $\pi/7$ (white or gray triangle in Figs.~\ref{fig:7kagome-sites} and~\ref{fig:klein}), called a (2,3,7) {\it Schwarz triangle}. Each white triangle is related to a grey triangle by an orientation-reversing transformation. The tiling of the Poincar\'e disk by infinitely many (2,3,7) Schwarz triangles is described by the hyperbolic {\it triangle group} $\Delta(2,3,7)$, an infinite group which contains both orientation-preserving (conformal) and orientation-reversing (anti-conformal) transformations, and can be viewed as a space group. Combining one white and one grey triangle into a larger triangle (blue or green triangles in Fig.~\ref{fig:klein}), we obtain a tiling of the Klein quartic by 168 such {\it fundamental triangles}. The tiling of the entire Poincar\'e disk by infinitely many fundamental triangles is given by the {\it von Dyck group} $D(2,3,7)$, which is a subgroup of index 2 in $\Delta(2,3,7)$. The von Dyck group consists exclusively of conformal transformations. It is generated by the counter-clockwise rotations $e_1,e_2,e_3$ by $\pi$, $2\pi/3$, and $2\pi/7$, respectively, about the $\pi/2$, $\pi/3$, and $\pi/7$ vertices of one given Schwarz triangle:
\begin{align}
    \!\!D(2,3,7)=\langle e_1,e_2,e_3:e_1e_2e_3=e_1^2=e_2^3=e_3^7=\mathbbold{1}\rangle.\!
\end{align}
The Fuchsian translation group $\Gamma_{\{14,7\}}$ is a normal subgroup of index 168 in $D(2,3,7)$. The factor group:
\begin{align}\label{AutX}
    G=D(2,3,7)/\Gamma_{\{14,7\}}\cong\mathrm{PSL}(2,\mathbb{Z}_7),
\end{align}
is the order-168 group of conformal automorphisms of the Klein quartic. Each fundamental triangle is a fundamental domain for the action of this group.

This automorphism group can be used to systematically construct CLS as follows. Each vertex of the heptagon-kagome lattice (Fig.~\ref{fig:klein}, blue dot) is associated to a rhombus consisting of two fundamental triangles (cf.~Fig.~\ref{fig:7kagome-sites}) that are related to each other by a $\pi$ rotation about that vertex. 
Denote by $z_1$ an arbitrary initial site of the heptagon-kagome lattice, and by $\psi_1(z)$ the known heptagon-pair wave function associated with it. We can always choose the generators of $D(2,3,7)$ such that $e_1$ corresponds to the $\pi$ rotation about $z_1$. Denote by $G_{z_1}$ the {\it stabilizer subgroup} (or isotropy group) of $z_1$; this is the subgroup of $G$ that leaves $z_1$ invariant. It is generated by $e_1$ and isomorphic to $\mathbb{Z}_2$. It is of index $|G:G_{z_1}|=|G|/|G_{z_1}|=84$ in $G$; its 84 left cosets correspond to the 84 inequivalent rhombi in the Klein cell, and thus 84 inequivalent sites of the heptagon-kagome lattice. 
(In fact, they are precisely the rhombi of the heptagon-dice lattice studied in Sec.~\ref{sec:7dice}.)
Indeed, the set $\{z_i:i=1,\ldots,84\}$ of heptagon-kagome sites in the Klein cell is equal to the {\it orbit} of $z_1$ under $G$, $G(z_1)=\{g(z_1):g\in G\}$. By the {\it orbit-stabilizer theorem}, $G(z_1)$ is in one-to-one correspondence with the set $G/G_{z_1}$ of left cosets of $G_{z_1}$ in $G$. 
In GAP, we can directly compute the coset space $G/G_{z_1}$ and use its elements to label the 84 sites of the heptagon-lattice. By acting with all elements of $G$ on this coset space, we can construct from $\psi_1$ all the 83 remaining symmetry-related CLS, accounting for the correct PBC on the Klein cell.

\begin{figure}[t!]
\centering
    \includegraphics[width=\linewidth]{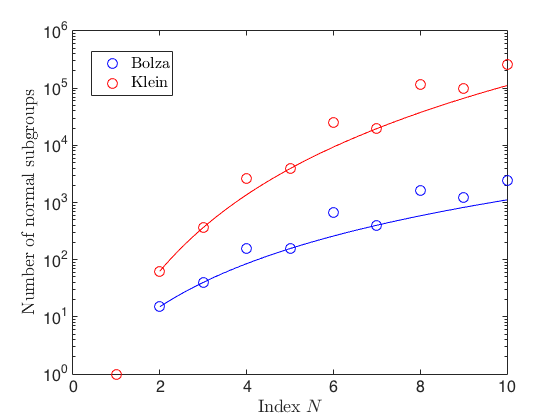}
    \caption[]{
    Number of normal subgroups of index $N$ in the Fuchsian translation groups $\Gamma_{\{8,8\}}$ (blue circles) and $\Gamma_{\{14,7\}}$ (red circles). The solid lines are the analytical result (\ref{NSGprime}) for subgroups of prime index.
    }
\label{fig:NSG}
\end{figure}

To be precise, one must account for TBC on the Klein cell, which involves $\b{k}$-dependent Bloch phase factors that must be determined. To compute these phase factors, we must distinguish between the CLS wave functions $\psi_i(z)$ in the extended Poincar\'e disk, and those on the compactified Klein cell, $\tilde{\psi}_i(z)$:
\begin{align}
    \psi_i(z)=\sum_{j=1}^{12}(-1)^j\delta_{z,z_{ij}},\hspace{5mm}
    \tilde{\psi}_i(z)=\sum_{j=1}^{12}(-1)^j\delta_{z,\tilde{z}_{ij}},
\end{align}
where the sum ranges over all 12 sites of the heptagon-pair state with nonzero support (see Fig.~\ref{fig:7kagome-CLS}), $z_{ij}$ denotes the coordinate of site $j$ in CLS $i$ in the extended Poincar\'e disk, $\tilde{z}_{ij}$ denotes that same coordinate inside the compactified Klein cell, and $\delta_{z_1,z_2}$ is the Kronecker symbol. 
To determine the $\b{k}$-dependent phase factors, for each $i=1,\ldots,84$ and $j=1,\ldots,12$ we must find the Fuchsian group element $\gamma_{ij}\in\Gamma_{\{14,7\}}$ such that $z_{ij}=\gamma_{ij}\tilde{z}_{ij}$, which can be done in GAP. This group element can then be expressed as a word in the 7 generators $\gamma_j$ in Eq.~(\ref{FuchsianKlein}). Upon substituting $\gamma_j\mapsto e^{ik_j}$ with $k_7$ given by Eq.~(\ref{k7}), we obtain a $\b{k}$-dependent phase factor, $e^{i\varphi_{ij}(\b{k})}$, which is then inserted in the wave function on the compactified Klein cell.
The final form of the 84 flat-band Bloch states is then:
\begin{align}\label{FBSBloch7kag}
    \Psi_{i,\b{k}}(z)=\sum_{j=1}^{12}e^{i\varphi_{ij}(\b{k})}(-1)^j\delta_{z,\tilde{z}_{ij}}.
\end{align}
Finally, we verify analytically that all such states are eigenstates of the Bloch Hamiltonian $H(\b{k})$ with energy $E(\b{k})=-2t$, and that out of them, only 28 are linearly independent for arbitrary $\b{k}$.

\subsubsection{Non-Abelian states}\label{eqn:7-kagome-nA}

To study non-Abelian states, we must first construct normal subgroups of the Fuchsian translation group $\Gamma_{\{14,7\}}$ for the underlying $\{14,7\}$ Bravais lattice. In Ref.~\cite{Maciejko:2022}, only normal subgroups of $\Gamma_{\{8,8\}}$ were computed. Due to the larger number of generators, the number of normal subgroups grows faster than for the $\{8,8\}$ Bravais lattice, limiting our evaluation to $N\leq 10$ (Fig.~\ref{fig:NSG}). 
For a strictly hyperbolic genus-$g$ Fuchsian group $\Gamma$, which is isomorphic to the fundamental group of a genus-$g$ surface,  $\Gamma\cong\pi_1(\Sigma_g)$~\cite{Katok}, the number of normal subgroups of prime index $N=p$ grows polynomially~\cite{Maciejko:2022}:
\begin{align}\label{NSGprime}
    \mathrm{NSG}_p=\frac{p^{2g}-1}{p-1}=1+\cdots+p^{2g-1},
\end{align}
where $g=2$ for $\Gamma_{\{8,8\}}$ and $g=3$ for $\Gamma_{\{14,7\}}$. For nonprime values of $N$, which are required to obtain non-Abelian PBC clusters, we perform the subgroup enumeration using GAP~\cite{LINS}.

While for the $\{8,8\}$ Bravais lattice, it was shown~\cite{Maciejko:2022} that non-Abelian clusters do not appear for values of $N$ less than 12, we find that for the $\{14,7\}$ Bravais lattice, there are non-Abelian clusters already at $N=6,8\leq 10$.
We restrict ourselves to PBC clusters that form a connected region before PBC identification, and for which words of length at most 1 exist in the transversal for $\Gamma_\text{PBC}$ in $\Gamma_{\{14,7\}}$~\cite{Maciejko:2022}. 
Using projector matrices as before to project out the Abelian states, we find that for all non-Abelian clusters with $N=6$ and $N=8$, Eq.~(\ref{FBfrac2}) is satisfied with $f=1/3$. 

In Fig.~\ref{fig:7kagome-DoS}(a), we plot the exact spectrum for all the identified PBC clusters with $N=8$ Klein cells (672 sites of the heptagon-kagome lattice), as well as the Abelian HBT spectrum. 
In agreement with both the real-space and momentum-space analyses, exact diagonalization reveals that the flat band is separated from the remainder of the spectrum by a gap, confirming the band-touching index $w=0$. 
As for the cases previously discussed, the Abelian HBT spectrum provides a remarkably good approximation to the full spectrum of non-Abelian clusters, despite the presence of higher-dimensional irreps in the latter.

\section{Heptagon-dice lattice}
\label{sec:7dice}

\begin{figure}[t!]
\centering
    \includegraphics[width=\linewidth]{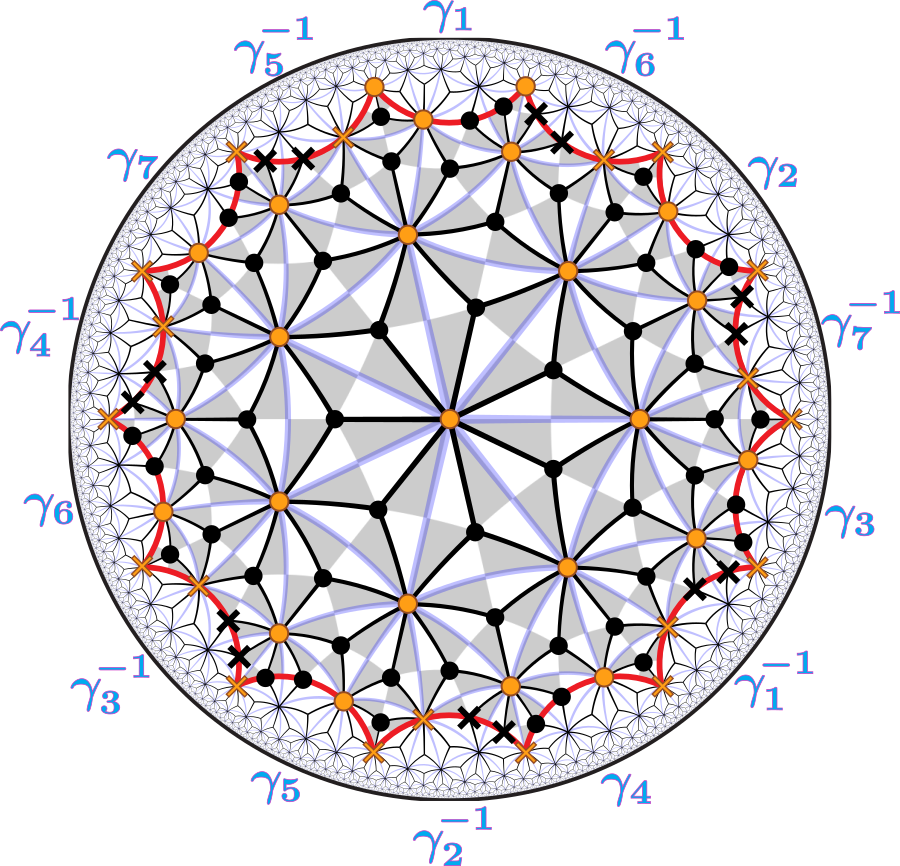}
    \caption[]{
    Heptagon-dice lattice. The Klein cell and the Fuchsian group generators are as in Fig.~\ref{fig:7kagome-sites}, but the cell now contains 80 inequivalent sites. 
    The lattice is bipartite, with 56 three-coordinated sites (black dots) and 24 seven-coordinated sites (orange dots) per Klein cell.
    }
\label{fig:7dice-sites}
\end{figure}

We finally turn to the last lattice considered in this study, the heptagon-dice lattice (Fig.~\ref{fig:7dice-sites}). Like the octagon-dice lattice (Sec.~\ref{sec:8dice}), it is a non-Euclidean generalization of the dice lattice, but this time with heptagonal symmetry. The heptagon-dice lattice is again bipartite, with hopping exclusively between sublattice $A$, consisting of three-coordinate sites, and sublattice $B$, consisting of seven-coordinated sites (black and orange dots in Fig.~\ref{fig:7dice-sites}, respectively). Each Klein cell contains 56 sites of the $A$ sublattice and 24 sites of the $B$ sublattice, for a total of 80 sites. A PBC cluster with $N$ Klein cells contains $N_A=56N$ and $N_B=24N$ sites of each sublattice, with a total of $N_A+N_B=80N$ sites. The bound in Eq.~(\ref{chiralFBS}) derived earlier implies $f\geq 32/80=2/5$ for the fraction of flat-band states at $E=0$, and $w\geq 0$ for the band-touching index. To pin down the exact values of $f$ and $w$, we analyze the hopping problem from both the real-space and reciprocal-space perspectives.

\subsection{Flat bands from real-space topology}
\label{sec:7dice-CLS}

\begin{figure}[t!]
\centering
    \includegraphics[width=0.855\linewidth]{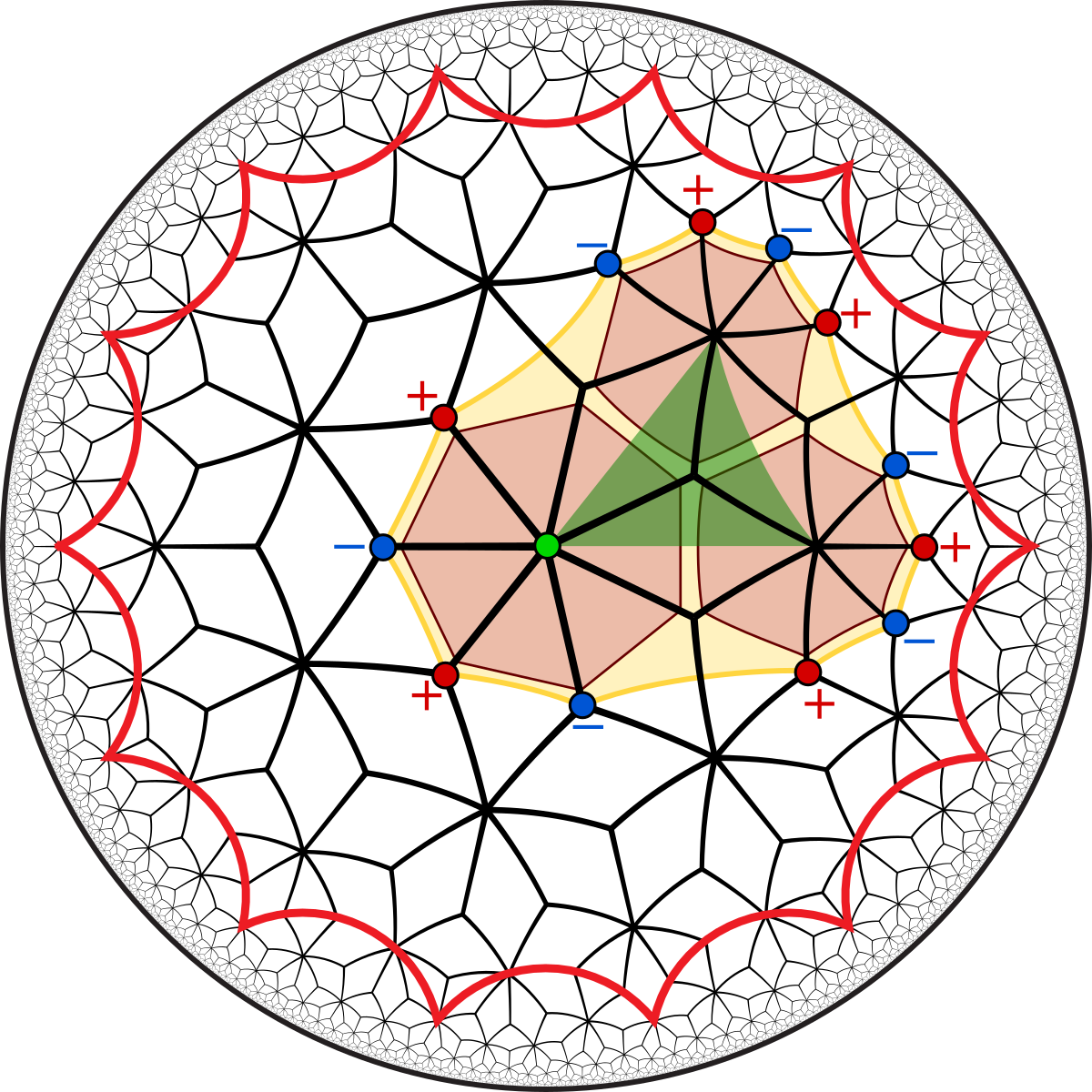}
    \caption[]{
    The maximally localized CLS state on the heptagon-dice lattice is the star-triplet state, bound to three heptagonal faces of the underlying $\{7,3\}$ lattice, shown in red. Equivalently, the CLS is bound to a single triangular face of the $\{3,7\}$ lattice (green triangle). 
    Red/blue dots denote probability amplitudes of alternating sign $\pm 1$. 
    An equal-weight superposition of the star-triplet states bound to the 7 triangles adjacent to a common seven-coordinated vertex (green dot) has cancelling amplitudes, implying a linear dependence relation. 
    }
\label{fig:7dice_CLS}
\end{figure}

We first study the spatially localized flat-band states. 
Based on the general discussion of chiral flat bands at the beginning of Sec.~\ref{sec:8dice}, we expect CLS to have support on the majority ($A$) sublattice only. 
In contrast with the Euclidean dice and the octagon-dice lattices, because the smallest polygon on the $A$ sublattice is an odd-sided heptagon, a CLS with probability amplitudes alternating in sign on a single heptagon is impossible. 
Instead, we find that the maximally localized CLS on the heptagon-dice lattice is the {\it star-triplet state} (Fig.~\ref{fig:7dice_CLS}). 
It is bound to a triplet of faces of the $\{7,3\}$ lattice, or equivalently, to a single triangular face of the $\{3,7\}$ lattice. 
There are 56 such triangular faces inside one Klein cell, thus there are 56 star-triplet states per Klein cell, leading to a total of $56N$ states on a PBC cluster with $N$ cells. 
However, the sum of the 7 star-triplet states associated with the 7 triangular faces around a seven-coordinate site (vertex of the $\{3,7\}$ lattice, indicated in green in Fig.~\ref{fig:7dice_CLS}) vanishes identically. 
Since there are 24 such vertices per Klein cell, or $24N$ vertices on a PBC cluster, we have $24N$ linear-dependence relations among the $56N$ star-triplet states, implying a total of $32N$ linearly independent CLS.

\begin{figure}[t!]
\centering
    \includegraphics[width=0.855\linewidth]{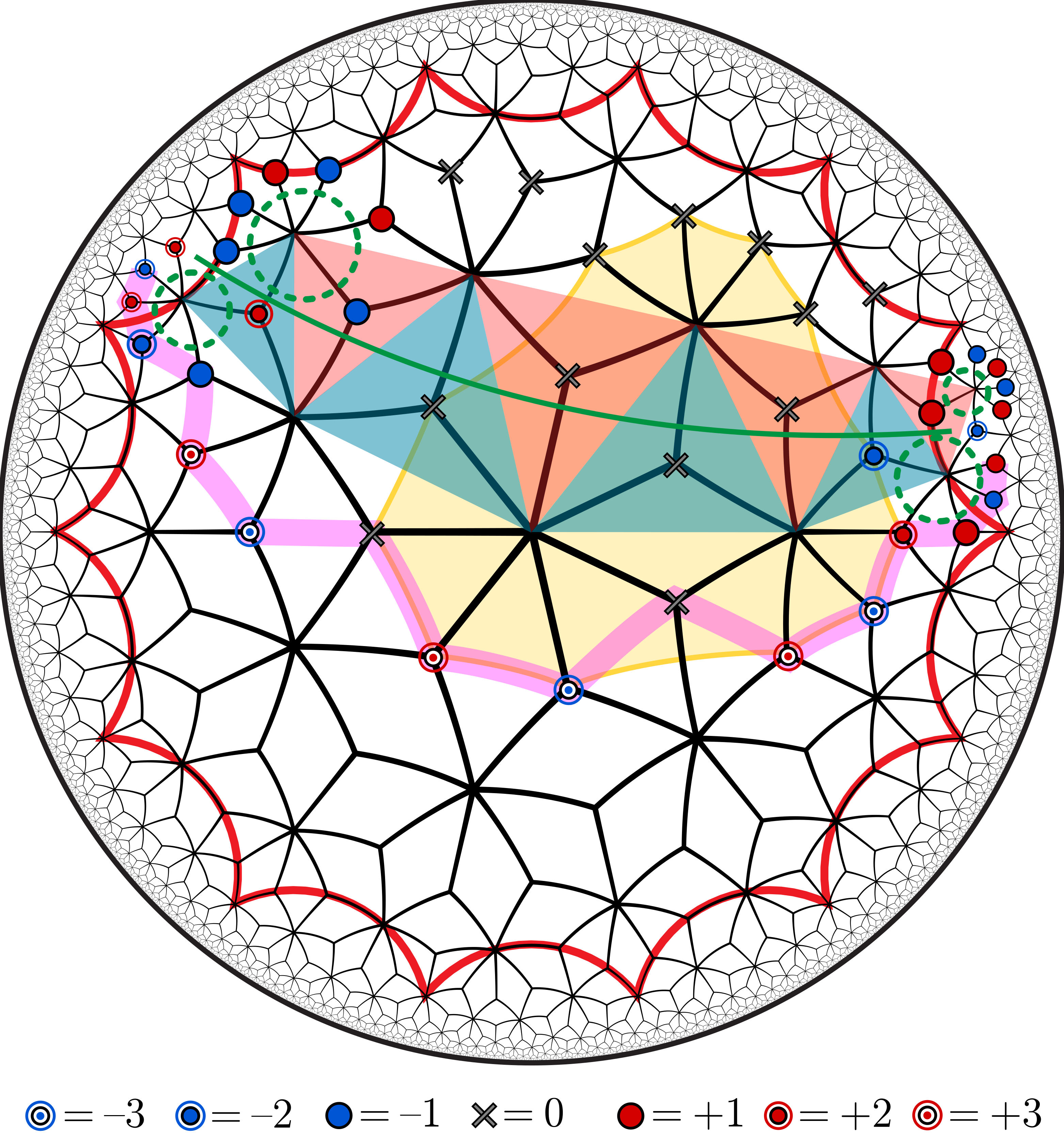}
    \caption[]{
    Construction of a flat-band noncontractible-loop state (with support on the magenta line) on the heptagon-dice lattice. 
    The state is obtained as a weighted superposition of the star-triplet CLS (Fig.~\ref{fig:7dice_CLS}) centered on the 8 highlighted triangles of the underlying $\{3,7\}$ lattice (the support of one such state is explicitly illustrated with the yellow outline). 
    The star-triplet CLS pertaining to the red (blue) triangles are assigned weight $+1$ ($+2$) in the superposition, and the resulting amplitudes on the individual sites are specified by seven distinct symbols according to the legend at the bottom of the figure.
    The state thus obtained for a finite string (green path) of star-triplet CLS on the Poincar\'e disk results in a chain CLS with a pair of heptagons attached at each end (green dashed circles). 
    Upon the Klein identification of the Klein cell boundary, the amplitudes on those heptagons interfere destructively and leave behind only the string that winds nontrivially around the Klein cell.
    }
\label{fig:7dice_homology}
\end{figure}

Next, we turn to noncontractible-loop states. In analogy with the heptagon-kagome lattice, we find that noncontractible-loop states can be obtained as linear superpositions of an extensive number of star-triplet CLS. To illustrate this, we depict in Fig.~\ref{fig:7dice_homology} the result of a superposition of 8 star-triplet CLS; each star-triplet CLS involved in the superposition is indicated by the $\{3,7\}$ triangle to which it is bound (cf.~Fig.~\ref{fig:7dice_CLS}). 
The color of the triangles represents the weight assigned to the corresponding CLS in the superposition; namely, the red (blue) triangle indicates weight $+1$ ($+2$). 
The result of this weighted superposition is a chain CLS whose amplitudes at the individual sites are represented by the seven species of symbols listed at the bottom of Fig.~\ref{fig:7dice_homology}. We find that the constructed chain CLS consists of a string (displayed in pink) along which the the wave-function amplitudes follow the repeating pattern ``$+3$, $-3$, $0$''. There are also residual nonzero amplitudes on a pair of heptagons (green dashed circles) appended at each end of the string. Note that by adding more star-triplet states, it is possible to separate out the two heptagon pairs arbitrarily far apart, and eventually annihilate them along a noncontractible loop of the $N$-cell PBC cluster.
In fact, for $N=1$ with the Klein identification, the heptagon pairs illustrated in Fig.~\ref{fig:7dice_homology} exactly cancel each other, leaving behind only the noncontractible-loop state.

Since the linearly-independent star-triplet CLS account for all flat-band states in the real-space approach (including in particular the noncontractible-loop states), we conclude that PBC clusters have $N_\text{FBS}=32N$ states at $E=0$, resulting in a flat-band fraction $f=2/5$ and a band-touching index $w=0$. 
As for the octagon-dice lattice, this saturates the chiral-asymmetry bound in Eq.~(\ref{chiralFBS}). 
Since $w=0$, we expect a finite gap separating the flat band from the rest of the spectrum, which is corroborated by numerical simulations [Fig.~\ref{fig:7dice-DoS}(a)].

\subsection{Flat bands from hyperbolic band theory}

In this section, we analyze the flat-band spectrum using HBT.

\subsubsection{Abelian states}
\label{sec:7dice_abelian}

\begin{figure}[t!]
\centering
    \includegraphics[width=0.99\linewidth]{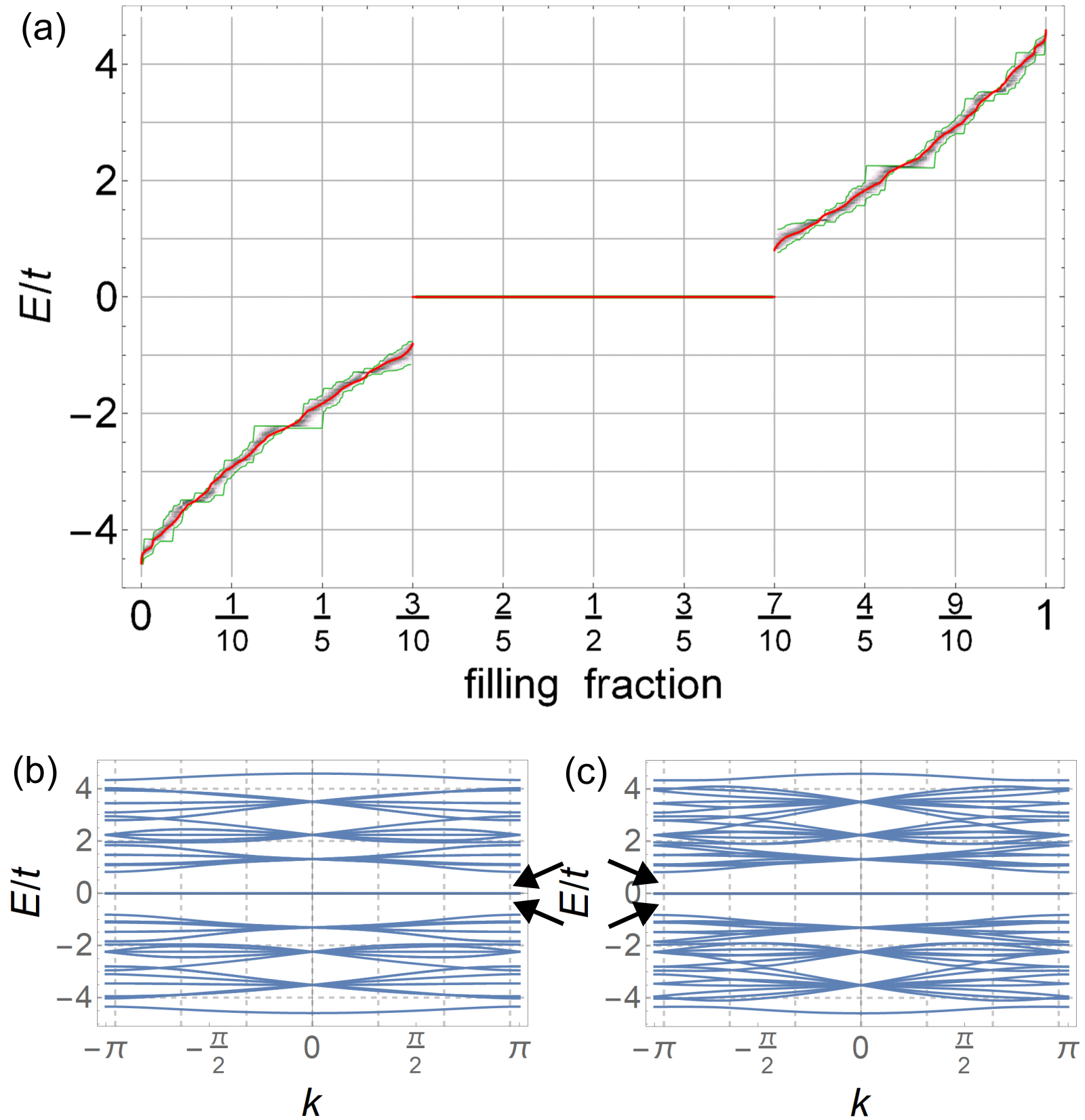}
    \caption[]{
    Spectrum of the heptagon-dice lattice ($V_B=0$).
    (a)~HBT spectrum (red) from $10^4$ momentum points in the 6D Brillouin zone vs.~exact spectrum (grayscale density plot bounded by the green range) from 27,995 PBC clusters with $N=8$ Klein cells (640 sites), of which 2,240 are non-Abelian.
    (b,c)~Abelian band dispersion along selected high-symmetry lines $\bs{k}=k\bs{n}$ in the 6D Brillouin zone. 
    The chosen directions are $\bs{n}=(1,0,0,0,0,0)$ in~(b), and $\bs{n}=(1,0,-1,0,1,0)$ in~(c). 
    Arrows indicate the energy gaps that separate the flat bands at $E=0$ from the dispersive bands at positive and negative energies.
    }
\label{fig:7dice-DoS}
\end{figure}

As for the heptagon-kagome lattice, the Brillouin zone for Abelian HBT states on the heptagon-dice lattice is 6D. Diagonalizing the $80\times 80$ Bloch Hamiltonian $H(\b{k})$, we find 32 flat bands at $E=0$, regardless of the value $V_B$ of the $B$ sublattice potential. Those 32 flat bands are found to be separated by an energy gap from the rest of the spectrum.
To visualize this gap, we show in Fig.~\ref{fig:7dice-DoS}(b,c) the Abelian band structure of the heptagon-dice lattice along selected lines in the 6D Brillouin zone.

To construct a Bloch basis for the 32 flat bands at generic momentum $\b{k}$, we proceed in analogy with Sec.~\ref{sec:7kagome_abelian} and exploit the symmetry of the Klein quartic. We begin with a single star-triplet state $\psi_1(z)$ with support on 12 sites entirely contained within one Klein cell, e.g., the CLS illustrated on Fig.~\ref{fig:7dice_CLS}. We wish to construct the remaining 55 star-triplet states that obey $\b{k}$-dependent TBC on the Klein cell. Once all 56 Bloch wave functions are thus constructed, we can check how many are linearly independent.

Let us describe the construction of the $56$ Bloch functions in more detail. Recall that for the heptagon-kagome lattice, we observed that each heptagon-pair CLS was associated to a rhombus in the (2,3,7) tiling of the Klein cell by its conformal automorphism group $G$. 
Those 84 rhombi in turn corresponded to 84 left cosets of the stabilizer subgroup $G_{z_1}\cong\mathbb{Z}_2\subset G$ of this rhombus. 
In the case of the heptagon-dice lattice, each star-triplet CLS is associated to an equilateral triangle of the $\{3,7\}$ lattice. 
Such a triangle consists of 6 Schwarz triangles of the triangle group $\Delta(2,3,7)$ (gray and white triangles in Fig.~\ref{fig:klein}). 
Equivalently, it consists of 3 fundamental triangles of the von Dyck group $D(2,3,7)$ or of its quotient group $G$ (green triangles in Fig.~\ref{fig:klein}). 
Denoting by $y_1$ the 3-coordinated site at the center of the initially chosen star-triplet state $\psi_1(y)$, its stabilizer subgroup $G_{y_1}$ is generated by the $2\pi/3$ rotation $e_2\in D(2,3,7)$ and is isomorphic to $\mathbb{Z}_3$. 
It is of index $|G:G_{y_1}|=|G|/|G_{y_1}|=56$ in $G$.
As before, we use the orbit-stabilizer theorem to equate the 56 orbits $G(y_1)$ with the left cosets $G/G_{y_1}$, which are computed in GAP. 
Using GAP also to determine the correct $\b{k}$-dependent Bloch phase factors as explained in Sec.~\ref{sec:7kagome_abelian}, the 56 flat-band Bloch wave functions can be written again in the form of Eq.~(\ref{FBSBloch7kag}). 
Using a random sampling of $10^4$ momentum points in the 6D Brillouin zone, we confirm that the rank of the eigenvector matrix is always 32. The Abelian HBT analysis thus again predicts $f=2/5$ and $w=0$, in accordance with Eq.~(\ref{FBfrac1}).

\subsubsection{Non-Abelian states}

Following the same steps as discussed in Sec.~\ref{eqn:7-kagome-nA}, we use exact numerical diagonalization in real space to obtain the spectrum of PBC clusters of the heptagon-dice lattice with up to $N=8$ Klein cells (640 sites). The results for $N=8$ are plotted in Fig.~\ref{fig:7dice-DoS}(a) for $V_B=0$. We further find that the two observed energy gaps remain (but become asymmetric) for $V_B\neq 0$.
Again, we observe that the Abelian HBT spectrum approximates well the exact spectrum of non-Abelian PBC clusters. 
Projecting out the Abelian states, we find that the non-Abelian spectrum obeys Eq.~(\ref{FBfrac2}) with the same value of $f=2/5$.

\section{Summary and outlook}
\label{sec:summary}

In conclusion, we have performed a study of flat bands and band-touching phenomena in hyperbolic lattices, using a combination of analytical and numerical techniques. 
We have focused on nearest-neighbor hopping models defined on four distinct hyperbolic lattices: the octagon-kagome and heptagon-kagome lattices, which generalize the Euclidean kagome lattice, and the octagon-dice and heptagon-dice lattices, which generalize the Euclidean dice (also called rhombille) lattice. For all four lattices, we have calculated the flat-band fraction $f$ and the band-touching index $w$. The former characterizes the extensive (linear) growth of the flat-band degeneracy with system size, while the latter signals the existence of an additional degeneracy at the flat-band energy that remains finite in the thermodynamic limit.

First, we found that these quantities could be obtained from real-space topology arguments involving a careful counting of linearly independent CLS and noncontractible-loop states, as in Ref.~\onlinecite{Bergman:2008}, but with suitable generalization to hyperbolic geometry. 
Key ingredients in this real-space topology approach were the partition of a hyperbolic lattice into Bravais unit cells related by non-Euclidean translations, which form a non-Abelian (Fuchsian) translation group, as well as the proper formulation of PBC for hyperbolic lattices. 
In particular, the negative curvature of hyperbolic lattices implies that PBC clusters with $N$ cells are compactified on a surface whose genus grows linearly with $N$ [Eq.~(\ref{eqn:Bolza-cluster-genus})], which suggests a crucial role played by flat-band states with support on noncontractible loops winding around the cluster.
While for the lattices with odd-sided polygons (heptagon-kagome and heptagon-dice) such noncontractible-loop states are shown to be linear combinations of CLS, they provide an indispensable contribution to the flat-band degeneracy for the lattices with even-sided polygons (octagon-kagome and octagon-dice). In the latter case, noncontractible-loop states are also responsible for the nonzero band touching index $w\neq 0$, implying that these band touchings are protected by real-space topology.

Second, we analyzed the spectrum of the four chosen hyperbolic lattices from a reciprocal-space perspective, using hyperbolic band theory as well as numerical diagonalization. 
We separately studied Abelian Bloch states, which transform according to 1D irreps of the Fuchsian group, and non-Abelian Bloch states, which transform according to higher-dimensional irreps. We found that the index $w$ precisely counts the number of touching points between the flat band and dispersive bands in the \emph{Abelian} band structure, hence its interpretation as a band-touching index. 
Using numerical exact diagonalization on finite lattices with several hundred sites, we also found that both Abelian and non-Abelian Bloch states are characterized by the \emph{same} flat-band fraction $f$. 
To our understanding, this is the first concrete characterization of non-Abelian Bloch states. 
Our results above lead us to conjecture that the following properties hold for \emph{all} flat-band hyperbolic lattices: (1) the flat-band fraction $f$ is equal for Abelian and non-Abelian states, and (2) the $w$ additional states that lead to band touching are Abelian.

We note in this context that Ref.~\onlinecite{saa2021} uses graph-theoretic methods to predict the value of $f$ for the line graph of a generic $\{p,q\}$ lattice, which evaluates to $f=1/3$ for $\{p,3\}$ lattices with PBC, in agreement with our results for the kagome-like lattices. 
In our work, we further show how this number can be obtained from real-space topology and hyperbolic band theory arguments, and that it applies separately to both Abelian and non-Abelian states. We also investigate band-touching phenomena which are not explicitly discussed for the heptagon-kagome and octagon-kagome lattices in Ref.~\onlinecite{saa2021}. 
For the dice-like lattices, our results suggest that the lower bound $f\geq|N_A-N_B|/(N_A+N_B)$ implied by Eq.~(\ref{chiralFBS}), which evaluates to $f\geq(p-3)/(p+3)$ for dice decorations of the $\{p,3\}$ lattice, is generically saturated.

Looking ahead, we propose a few directions for future research. First, it would be desirable to rigorously prove or disprove the conjectured properties (1) and (2) above. (If the band-touching index is $w=1$, property (2) is trivially true, since non-Abelian states necessarily appear in degenerate multiplets.) Second, the hyperbolic band theory analysis presented here requires the determination of a Fuchsian translation group for a given hyperbolic lattice. 
To date, this has only been achieved by Ref.~\onlinecite{Boettcher:2021} for a limited set of $\{p,q\}$ lattices. For example, it would be desirable to identify the hyperbolic Bravais lattice for the $\{9,3\}$ lattice, whose line graph is known to exhibit a gapped flat band~\cite{Kollar:2019}, and which also provides the skeleton for the nine-fold symmetric version of the dice lattice. In addition, our understanding of the spectra of hyperbolic lattices and of non-Abelian Bloch states would be greatly advanced if an explicit parametrization of the moduli space of irreps of two and higher dimensions could be found.

Another promising direction suggested by the present work is to construct hyperbolic lattice models with nearly-flat topological bands. 
By analogy with the Euclidean kagome lattice~\cite{ohgushi2000}, we anticipate nearly-flat bands with nonzero Chern number in the octagon-kagome lattice with magnetic fluxes threading the triangular and octagonal plaquettes of this lattice.
Placing correlated fermions or bosons in such topologically nontrivial nearly-flat bands may result in exotic fractional Chern insulator states, as in the Euclidean case~\cite{tang2011,sun2011,Neupert:2011,sheng2011,regnault2011,parameswaran2013}. 
Owing to the underlying negative curvature however, such states could harbor exotic features with no direct Euclidean analog. 
For example, the Riemann-Hurwitz formula implies that the genus of compactified PBC clusters grows linearly with the system size, a fact that was central for the real-space topological argument presented in this work. 
This in turn suggests that putative fractional Chern insulator states on a hyperbolic lattice possess a topological ground-state degeneracy that is exponentially large in the system size~\cite{wen1990}, an effect that is somewhat reminiscent of the phenomenology of fracton phases of matter~\cite{Nandkishore:2019}.

Finally, we hope this work will stimulate further experimental progress in the field of hyperbolic lattices. 
It would be interesting to exploit the platforms of Refs.~\onlinecite{Kollar:2019,Lenggenhager:2021} to perform explicit experimental observations of CLS and noncontractible-loop states. 
This has been done for Euclidean photonic lattices with flat bands in Refs.~\onlinecite{vicencio2015,mukherjee2015,zong2016,xia2018,ma2020}.
Implementations of nearly-flat topological bands via magnetic flux threading in the octagon-kagome lattice would also be a desirable goal. 
In the classical circuit-network architecture, complex-phase electric circuit elements~\cite{Hofmann:2019,chen2022} could be combined with kagome-like versions of the hyperbolic circuits of Refs.~\onlinecite{Lenggenhager:2021,zhang2022}. 
In the quantum context, a protocol for engineering plaquette fluxes for photons in a microwave cavity array was demonstrated in Ref.~\onlinecite{owens2018}, and could in principle be utilized with cQED hyperbolic lattices~\cite{Kollar:2019}.
If additionally coupled to superconducting qubits~\cite{carusotto2020,bienias2022}, such photons would be governed by a strongly interacting Bose-Hubbard model~\cite{ma2019} with topological hyperbolic bands that could potentially support exotic correlated ground states.

\section*{Code and data availability}

All the code (written in \textsc{Wolfram Mathematica}, \textsc{Matlab}, and \textsc{GAP}) as well as the generated data used to arrive at the conclusions presented in this work are publicly available in the data repository~\cite{supp}.

\acknowledgements

We acknowledge helpful discussions with I.~Boettcher, A.~Chen, C.~Doran, P.~M.~Lenggenhager, T.~Neupert, M.~Protter, S.~Rayan, G.~Shankar, S.~Stadnicki, A.~Stegmaier, R.~Thomale, L.~Upreti, and D.~M.~Urwyler. T.~B.~was supported by the Ambizione grant No.~185806 by the Swiss National Science Foundation.
J.M.~was supported by NSERC Discovery Grants \#RGPIN-2020-06999 and \#RGPAS-2020-00064; the Canada Research Chair (CRC) Program; the Government of Alberta's Major Innovation Fund (MIF); the Tri-Agency New Frontiers in Research Fund (NFRF, Exploration Stream); and the Pacific Institute for the Mathematical Sciences (PIMS) Collaborative Research Group program.
The basic layout for illustrations involving kagome lattices was generated with the help of the \texttt{Tess} package~\cite{tess} in \textsc{Wolfram Mathematica}, and that of the dice lattices by adapting Wikipedia public domain illustrations generated by user Parcly Taxel~\cite{wiki}.

\end{document}